\documentclass[conference]{IEEEtran}
\usepackage{adjustbox}
\usepackage{multirow}
\usepackage{multicol}
\usepackage[cmex10]{amsmath}
\usepackage{amssymb}
\usepackage[tight,footnotesize]{subfigure}
\usepackage{cite}
\usepackage{algorithmic}
\usepackage{algorithm}
\usepackage{url}
\usepackage{threeparttable}
\usepackage{mathrsfs}
\usepackage{amsfonts}
\usepackage{graphicx}
\usepackage{bbding}
\usepackage{amsthm}
\usepackage{color}
\usepackage{todonotes}

\newtheorem{Theorem}{Theorem}
\newtheorem{Definition}{Definition}

\begin{document}
%
\title{The Power of Bamboo: On the Post-Compromise Security for Searchable Symmetric Encryption}

\author{\IEEEauthorblockN{Tianyang Chen\IEEEauthorrefmark{1}$^,$\IEEEauthorrefmark{2}$^,$\IEEEauthorrefmark{4},
Peng Xu\Envelope$^,$\IEEEauthorrefmark{1}$^,$\IEEEauthorrefmark{2}$^,$\IEEEauthorrefmark{4},
Stjepan Picek\IEEEauthorrefmark{8},
Bo Luo\IEEEauthorrefmark{6},
Willy Susilo\IEEEauthorrefmark{7},
Hai Jin\IEEEauthorrefmark{1}$^,$\IEEEauthorrefmark{3}$^,$\IEEEauthorrefmark{4},
Kaitai Liang\IEEEauthorrefmark{5}}
\IEEEauthorblockA{\IEEEauthorrefmark{1}National Engineering Research Center for Big Data Technology and System, Services Computing Technology and System Lab}
\IEEEauthorblockA{\IEEEauthorrefmark{2}Hubei Key Laboratory of Distributed System Security,\\
Hubei Engineering Research Center on Big Data Security, School of Cyber Science and Engineering}
\IEEEauthorblockA{\IEEEauthorrefmark{3}Cluster and Grid Computing Lab, School of Computer Science and Technology}
\IEEEauthorblockA{\IEEEauthorrefmark{4}Huazhong University of Science and Technology, Wuhan, 430074, China}
\IEEEauthorblockA{\IEEEauthorrefmark{8}Digital Security Group, Radboud University, Nijmegen, The Netherlands}
\IEEEauthorblockA{\IEEEauthorrefmark{6}Department of EECS and Institute of Information Sciences, The University of Kansas, Lawrence, KS, USA}
\IEEEauthorblockA{\IEEEauthorrefmark{7}Institute of Cybersecurity and Cryptology, School of Computing and Information Technology,\\
University of Wollongong, Wollongong, NSW 2522, Australia}
\IEEEauthorblockA{\IEEEauthorrefmark{5}Faculty of Electrical Engineering, Mathematics and Computer Science, \\
Delft University of Technology, 2628 CD Delft, The Netherlands}
\IEEEauthorblockA{\{chentianyang, xupeng\}@mail.hust.edu.cn, stjepan.picek@ru.nl, bluo@ku.edu,\\
wsusilo@uow.edu.au, hjin@hust.edu.cn, Kaitai.Liang@tudelft.nl}}

\IEEEoverridecommandlockouts
\makeatletter\def\@IEEEpubidpullup{6.5\baselineskip}\makeatother

\maketitle


\begin{abstract}
Dynamic searchable symmetric encryption (DSSE) enables users to delegate the keyword search over dynamically updated encrypted databases to an honest-but-curious server without losing keyword privacy.
This paper studies a new and practical security risk to DSSE, namely, \emph{secret key compromise} (e.g., a user's secret key is leaked or stolen), which threatens all the security guarantees offered by existing DSSE schemes.
To address this open problem, we introduce the notion of \emph{searchable encryption with key-update} (SEKU) that provides users with the option of non-interactive key updates.
We further define the notion of post-compromise secure with respect to leakage functions to study whether DSSE schemes can still provide data security after the client's secret key is compromised. We demonstrate that post-compromise security is achievable with a proposed protocol called ``\texttt{Bamboo}".
Interestingly, the leakage functions of \texttt{Bamboo} satisfy the requirements for both forward and backward security.
We conduct a performance evaluation of \texttt{Bamboo} using a real-world dataset and compare its runtime efficiency with the existing forward-and-backward secure DSSE schemes.
The result shows that \texttt{Bamboo} provides strong security with better or comparable performance.
\end{abstract}

\section{Introduction}\label{SEC.INTRODUCTION}
\footnotetext{\Envelope Peng Xu is the corresponding author.}
\subsection{Motivation}

Dynamic searchable symmetric encryption (DSSE)~\cite{DBLP:conf/ccs/KamaraPR12}, a type of structured encryption~\cite{DBLP:conf/asiacrypt/ChaseK10, DBLP:conf/crypto/KamaraMO18,DBLP:conf/eurocrypt/GeorgeKM21}, is a cryptographic tool that enables a client to outsource its encrypted database (i.e., a collection of ciphertexts) to a remote server and further perform the dynamic data update and keyword search over the encrypted database, where the server is modeled as an honest-but-curious adversary.
The client is allowed to maintain a secret key locally to generate secure keyword search queries and to issue data update queries to the server for adding/deleting ciphertexts to/from the encrypted database.
DSSE provides secure queries over encrypted cloud-based databases and has been widely deployed in real-world applications, such as Lookout~\cite{Lookout}, bitglass~\cite{bitglass}, Cossack Labs' Acra~\cite{cossacklabs}, and MVISION Cloud~\cite{mvision}.

Since the seminal work of DSSE by Kamara et al.~\cite{DBLP:conf/ccs/KamaraPR12}, many research efforts have been devoted to constructing practical and efficient DSSE schemes.
To provide high efficiency on data update and keyword search, DSSE discloses some information to the server, causing potential information leakage~\cite{DBLP:conf/ccs/CurtmolaGKO06}.
A well-studied solution to minimizing such information leakage is to use forward security~\cite{DBLP:conf/ndss/StefanovPS14,DBLP:conf/ccs/Bost16} to defend against attacks such as file-injection attack~\cite{DBLP:conf/uss/ZhangKP16}.
The idea behind forward security is to prevent data update queries from leaking the updated keywords.
Bost et al. introduced another security notion, backward security~\cite{DBLP:conf/ccs/BostMO17}, which limits the information leaked from the deleted ciphertexts during search queries.
Since then, various DSSE schemes have been proposed to achieve both forward and backward security without loss of efficiency, e.g.,~\cite{DBLP:conf/ccs/ChamaniPPJ18,DBLP:conf/ccs/SunYLSSVN18,DBLP:conf/ndss/SunSLYSLNG21,DBLP:conf/ndss/DemertzisCPP20,DBLP:conf/esorics/ZuoSLSP19, DBLP:conf/uss/ChamaniPMD22,DBLP:conf/esorics/ChenXWZSJ21,9724186}.

The security of all existing DSSE schemes relies on a ``strong" assumption that the secret key of the client can always be protected and will not be compromised. Unfortunately, this assumption may not scale well in practice.
For example, 23,000 secret keys of HTTPS certificates were compromised by network attackers~\cite{trustico-key-leak}, and Imperva inc. leaked customers' data after its cloud API key was stolen~\cite{imperva-key-leak}.
In the context of DSSE, if a client's secret key is exposed, the thief who obtains the key can easily break the encrypted database and observe data updates and search queries.
In this paper, we pose the following question:
\emph{``Can a DSSE scheme still provide clients with data security while maintaining its performance efficiency if the client's secret key is compromised?"}

Due to the requirement of maintaining high performance, it may be infeasible to design a perfectly secure DSSE system that can protect a client's secret key at all times against various attacks.
Hence we aim to increase the difficulty of attacking and limit the information leakage if the key is compromised.
Our solution adds new features to DSSE that enable the clients to update the secret key based on their preferences.
We would like to mention that this philosophy complies with the recommendations by the Data Security Standard of Payment Card Industry~\cite{PCIDSS} and NIST Special Publication 800-57~\cite{NIST800-57}.

\begin{table*}[h]
\centering
\caption{Comparisons of \texttt{Bamboo} with Related Forward-and-Backward Secure DSSE Schemes.}\label{T.Comparsion0}
\begin{threeparttable}
\begin{tabular}{|c|c|c|c|c|c|c|c|c|c|}
\hline \multirow{2}{*}{Scheme}  & \multicolumn{3}{c|}{\textsf{KeyUpdate}} &\multicolumn{2}{c|}{Search} & Data Update & Client\\
\cline{2-7}  &  Server Computation & Client Computation & Bandwidth &Computation &Bandwidth &Computation \& Bandwidth & Storage\\
\hline \texttt{Fides}~\cite{DBLP:conf/ccs/BostMO17}  & $O(1)$ & $O(N)$ & $O(N)$ &$O(a_w)$ &$O(a_w)$  &$O(1)$ & $O(W\text{log}F)$ \\
\hline \texttt{Aura}~\cite{DBLP:conf/ndss/SunSLYSLNG21} & $O(1)$ & $O(N)$ & $O(N)$ &$O(n_w)$ &$O(n_w)$ &$O(1)$ & $O(Wd_{\text{max}})$ \\
\hline \texttt{SD}$_a$~\cite{DBLP:conf/ndss/DemertzisCPP20} & $O(1)$ & $O(N)$ & $O(N)$ &$O(a_w + \text{log}N)$ &$O(a_w+\text{log}N)$ &$O(\text{log}N)$  & $O(1)$ \\
\hline \texttt{SD}$_d$~\cite{DBLP:conf/ndss/DemertzisCPP20}  & $O(1)$ & $O(N)$ & $O(N)$ &$O(a_w + \text{log}N)$ &$O(a_w+\text{log}N)$ &$O(\text{log}^3N)$ & $O(1)$ \\
\hline \texttt{MITRA}~\cite{DBLP:conf/ccs/ChamaniPPJ18}  & $O(N)$ & $O(1)$ & $O(1)$ &$O(a_w)$ &$O(a_w)$ &$O(1)$ & $O(W\text{log}F)$ \\
\hline
\hline \textbf{\texttt{Bamboo}} (Ours) & $O(N)$ & $O(1)$& $O(1)$ &$O(a_w)$ &$O(a_\text{max})$ &$O(1)$  & $O(W\text{log}F)$\\
\hline
\end{tabular}
\begin{tablenotes}[flushleft]
\item The \textsf{KeyUpdate} costs of \texttt{Fides}, \texttt{Aura}, $\texttt{SD}_a$, and $\texttt{SD}_d$ are counted from the trivially implemented \textsf{KeyUpdate}, and the \textsf{KeyUpdate} of \texttt{MITRA} is implemented by replacing the original PRF with the key-updatable PRF~\cite{9724186}. $N$ denotes the total number of keyword-and-file-identifier pairs, $W$ denotes the number of distinct keywords, $F$ denotes the number of files, and $d_{\text{max}}$ denotes the allowed maximum number of deletion queries. For keyword $w$, $a_\text{max}$ is the padding constant used for hiding the real search result size, $a_w$ is the total number of data update queries the client has issued, and $n_w$ is the number of files containing $w$. All the listed schemes achieve $O(N)$ storage complexity on the server side.
\end{tablenotes}
\end{threeparttable}
\end{table*}

\subsection{Simple Solutions Do Not Work}\label{Sec.TravialSolution}
To implement our design idea, we equip DSSE with a new and secure protocol (which we call \textsf{KeyUpdate}) to update the secret key of the encrypted database.
A straightforward implementation of \textsf{KeyUpdate} is to mandate the client to download the entire encrypted database, decrypt the database, then re-encrypt it with a new secret key, and upload the re-encrypted database to the server.
Clearly, this trivial solution will lead to huge bandwidth and client computation costs, which are linear to the size of the encrypted database. For a large-scale database, this solution is impractical.
More importantly, it may fail to protect the security of ciphertexts generated in a ``special time slot": the period between the key being compromised and the time when the new key is updated.
Therefore, we require the proposed \textsf{KeyUpdate} protocol to deliver (1) efficiency: taking constant bandwidth and computational costs to delegate a secure \textsf{KeyUpdate} task to the server, and (2) security: ensuring the security of ciphertexts generated during the special time slot.

One potential approach is simply applying key-updatable tools to existing DSSE schemes for the \textsf{KeyUpdate} protocol with the desired efficiency and security.
Unfortunately, this is very unlikely, particularly to the security requirement of guaranteeing the security of ciphertexts that are generated with the compromised keys.
We take a close look at a pair of examples.
One may adopt \texttt{MITRA}~\cite{DBLP:conf/ccs/ChamaniPPJ18} to implement \textsf{KeyUpdate} by replacing the original PRF functions with those that enable key update, e.g.,~\cite{DBLP:conf/crypto/BonehLMR13,9724186}.
However, if the secret key is compromised, the thief (with the key) can easily learn the content of a newly updated ciphertext by traversing all the possible keywords and counters.
Similar vulnerability exists for \texttt{ORION} and \texttt{HORUS}~\cite{DBLP:conf/ccs/ChamaniPPJ18}.
One may also use \emph{ciphertext-independent updatable encryption}, e.g.,~\cite{DBLP:conf/eurocrypt/LehmannT18,DBLP:conf/eurocrypt/KloossLR19,DBLP:conf/crypto/BoydDGJ20}, to provide ORAM and OMAP~\cite{DBLP:conf/ccs/WangNLCSSH14} with key update function.
However, this extension cannot guarantee the security of ciphertexts generated in the special time slot mentioned earlier because the compromised secret key remains valid to decrypt the updatable ORAM and OMAP until it is updated. A more detailed discussion can be found in Section~\ref{Dis.TrivialExtToSEKU}.

\subsection{Ideas Behind \texttt{Bamboo}}\label{SEC.OURIDEA}

The proposed instance ``\texttt{Bamboo}'' implements both the traditional DSSE functions and an efficient \textsf{KeyUpdate} protocol that meets the two aforementioned requirements: efficiency and security. In this section, we explain how \texttt{Bamboo} protects the client's private data from being stolen while maintaining high search efficiency. The main ideas behind \texttt{Bamboo} are twofold: two-layer encryption and a hidden chain-like inter-ciphertext structure.

\texttt{Bamboo} uses a two-layer encryption mechanism to generate a ciphertext. The first layer (i.e., the inner layer) is used as a traditional encryption scheme, and the second one (i.e., the outer layer) is another encryption with the client's secret key designed for the key update.
To generate a ciphertext, \texttt{Bamboo} first chooses a random number as the encryption key for the first layer to encrypt the original data. Then the encrypted data will be encrypted again in the second layer with the client's secret key.

The two-layer encryption mechanism guarantees that, even when the thief compromises the client's secret key and all the historical random numbers, the thief still cannot reveal any information from a newly generated ciphertext $\mathbf{C}$. Although the thief can decrypt the second layer encryption of $\mathbf{C}$ with the compromised client's secret key, s/he cannot decrypt the first layer in the absence of the random number that was used to generate $\mathbf{C}$. This feature enables \texttt{Bamboo} to maintain the security of the ciphertexts generated during the special time slot.

The second layer helps \texttt{Bamboo} to perform a non-interactive \textsf{KeyUpdate} protocol. The non-interactive feature reduces the overhead in executing the \textsf{KeyUpdate} protocol. The key-updatable feature guarantees that the thief cannot leverage a historically compromised secret key to decrypt the ciphertexts generated after the secret key is updated because the new secret key is unknown to the thief.

To gain efficient search performance, \texttt{Bamboo} employs a hidden chain-like inter-ciphertext structure to organize ciphertexts encrypted by the same keyword. Specifically, for any two successively generated ciphertexts encrypted by the same keyword, the latter encrypts an index and the random number used in the former one. Upon receiving a search query, the server can find a matching ciphertext and decrypt the index and the random number in the next ciphertext matching the same keyword. In the same way, the server can find all matching ciphertexts. Hence, \texttt{Bamboo} achieves sub-linear search efficiency as most practical DSSE schemes do.

\subsection{Contributions}

\texttt{Bamboo} is one instance of a new type of DSSE, \emph{searchable encryption with key-update} (SEKU), that we introduced in this paper.
SEKU captures all the functionalities provided by traditional DSSE and supports the non-trivial \textsf{KeyUpdate} protocol.
We analyze the threat model, where the attacker could either be an honest-but-curious server or a malicious client who wants to steal the secret key of other clients and formalize post-compromise security via a common paradigm (i.e., leakage functions).
The security guarantees that neither the honest-but-curious server nor the malicious client can learn any semantic information from data update and \textsf{KeyUpdate} queries.
It subsumes forward security~\cite{DBLP:conf/ccs/Bost16} and is compatible with backward security~\cite{DBLP:conf/ccs/BostMO17}.
This means that a post-compromise secure SEKU instance can be both forward secure and backward secure.

We use the first SEKU scheme \texttt{Bamboo} to illustrate the above ideas and prove that \texttt{Bamboo} is post-compromise secure. The leakage functions of \texttt{Bamboo} satisfy the backward security requirements. \texttt{Bamboo} achieves an excellent balance between security and performance compared to the existing DSSE schemes that offer both forward security and the same level of backward security, as shown in Table I.
In terms of \textsf{KeyUpdate} complexity, only \texttt{Bamboo} and \texttt{MITRA} can delegate key update operations to the server and save a considerable amount of client computation and bandwidth costs compared to others, like \texttt{Fides}, \texttt{Aura}, $\texttt{SD}_d$, and $\texttt{SD}_a$.
For the search operation, \texttt{Bamboo} achieves the same level of computation as \texttt{Fides} and \texttt{MITRA} and outperforms $\texttt{SD}_a$ and $\texttt{SD}_d$.
The bandwidth cost of \texttt{Bamboo} in searching a keyword is dominated by a maximum padding value $a_\text{max}$ (defined by the client in practice).
$a_\text{max}$ helps \texttt{Bamboo} to achieve post-compromise security and we will elaborate in Section~\ref{SEC.IMPROVBAND} how to reduce the search bandwidth by adaptively adjusting the padding value.
\texttt{Bamboo} provides the same data update complexity as \texttt{Fides}, \texttt{Aura}, and \texttt{MITRA}, and better performance than $\texttt{SD}_a$ and $\texttt{SD}_d$.
Finally, \texttt{Bamboo} has the same storage complexity as the other schemes on the server side; the client storage cost is on par with \texttt{Fides} and \texttt{MITRA} and less than that in \texttt{Aura}, but a little higher than $\texttt{SD}_a$ and $\texttt{SD}_d$.
The experimental results show that the storage costs of \texttt{Bamboo} on both the server and client sides are practical (see Section \ref{SEC.EXPSTOR}).

We conduct a comprehensive evaluation of the performance of \texttt{Bamboo} using a real-world dataset (which is extracted from Wikipedia) and by comparing with \texttt{Fides}, \texttt{Aura}, and \texttt{MITRA}.
The results show that \texttt{Bamboo} outperforms the state-of-the-art schemes.
For example, when issuing a data update query, \texttt{Bamboo} saves 97.22\% and 68.33\% on client time costs compared to \texttt{Aura} and \texttt{Fides}, respectively.
When running \textsf{KeyUpdate} with 300 milliseconds network delay, \texttt{Bamboo} is 3.66 times and 2.57 times faster than \texttt{Fides} and \texttt{Aura}, respectively; and the client time cost is much smaller.
In terms of search performance, \texttt{Bamboo} outperforms \texttt{Fides} and significantly reduces client time costs as compared to both \texttt{Fides} and \texttt{MITRA}.

\section{Threat Model of SEKU}\label{SEC.SECMODEL}
SEKU should protect a client's encrypted database against two types of probabilistic polynomial time (PPT) adversaries.
One is widely recognized in the setting of DSSE, namely the honest-but-curious server, and the other, which is new to DSSE, is called the thief who steals the secret key.

{\bf Server.} Much like the traditional DSSE server, a SEKU server can store the encrypted database for a SEKU client, add/delete ciphertexts to/from the database, and further perform a secure keyword search.
%
Beyond that, SEKU enables the server to execute the key update, which is delegated by the client so that the secret key of the encrypted database can be updated to a new one.
%
The SEKU server here is honest while executing all predefined operations; meanwhile, it can observe the information leakage from those operations.

{\bf Thief.} We allow this party to compromise the client's secret key, eavesdrop on the communication between the client and the server, and obtain a copy of the encrypted database.
We do not assume that this party takes over the client or issues queries to the server with the compromised key.
We notice that some DSSE schemes may need the client to store extra secret information about the encrypted database (i.e., the private state) along with the secret key.
We here assume the secret information will also be leaked if the secret key is compromised.


SEKU provides two types of threat models based on whether there is collusion between the server and the thief (i.e., sharing their information).
The essential difference between these types is the adversarial ability of the server.
Specifically, collusion enables the server to learn the client's secret key so that it is no longer an honest-but-curious server but a fully malicious adversary.
This malicious server can reveal all the contents of the encrypted database with the key.
In this case, the security goal and the corresponding secure construction roadmaps will completely differ from the honest-but-curious case.

\begin{figure}[t]
\begin{minipage}[h]{\linewidth}
\centering
\includegraphics[width=\linewidth]{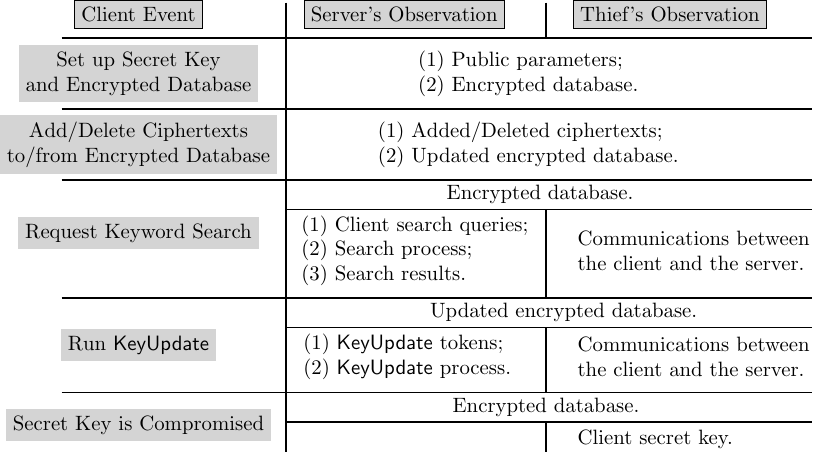}
\caption{The Observations on Client Queries and Key Compromise in the Moderate Threat Model. Note that in the stronger threat model, the two parties will share their observations.}\label{Fig.ThreatModel}
\end{minipage}
\end{figure}

\emph{Type 1: The Moderate Threat Model.}
The server and the thief are not allowed to collude.
The thief can be given the client's secret key and observe public parameters, data update queries, and the encrypted database.
While the client issues keyword search or \textsf{KeyUpdate} queries to the server, the thief can observe the communications between the client and the server.
Figure \ref{Fig.ThreatModel} outlines what the server and the thief can observe in this model.

\emph{Type 2: The Stronger Threat Model.}
Collusion between the two parties is allowed such that they can share exactly the same views as shown in Figure~\ref{Fig.ThreatModel}.
This makes the server so powerful that the following \emph{technical impossibilities} will incur. 

$\bullet$ \emph{Non-interactive \textsf{KeyUpdate} design.} In Type 2 threat model, the server is regarded as malicious and can use the compromised secret key of the client to break all the ciphertexts which were generated before the key compromise.
In this case, a secure \textsf{KeyUpdate} must hide the relationships between the pre-key-updated and post-key-updated ciphertexts from the server to avoid it leverages the knowledge of the pre-key-updated ciphertexts to infer information from the post-key-updated ones.
In other words, the \textsf{KeyUpdate} must update the encryption key of the encrypted database obliviously.
It is nearly impossible to design such a non-interactive \textsf{KeyUpdate}.
We notice that indistinguishability obfuscation ($i\mathcal{O}$)~\cite{DBLP:conf/stoc/JainLS21,DBLP:conf/eurocrypt/GayJLS21} could be a potential solution to offload the oblivious \textsf{KeyUpdate} to the server.
However, it is still unknown how to construct a practical $i\mathcal{O}$ scheme.
%
%
\\
\indent $\bullet$ \emph{Interactive \textsf{KeyUpdate} design.} The simple solution described in Section~\ref{Sec.TravialSolution} may achieve the oblivious \textsf{KeyUpdate}, but clearly, it is interactive and expensive.
Even if there may exist a practical and efficient interactive key update approach, there are still further concerns and impossibilities for the design.
Due to page limit, we leave the details in Section~\ref{Diss.DiscussType2Model}.

Because of these impossibilities, this work focuses on Type 1 model. 
We will define the security notions as the indistinguishability of one real and one simulated SEKU in Section~\ref{SEC.DEFINITION}, and formulate the leakage functions of post-compromise security in Section~\ref{SEC.LEAKAGEFUNC}.

\section{SEKU and Its Security Definitions}\label{SEC.DEFINITION}

\subsection{Notations}

Let $\mathcal{A}_\text{Srv}$ and $\mathcal{A}_\text{Thf}$ denote the server and the thief, respectively. Let $\lambda\in\mathbb{N}$ denote the security parameter.
We use $e\overset{\$}{\leftarrow}\mathcal{X}$ to denote uniformly sampling an element $e$ from a distribution or set $\mathcal{X}$.
For a set $\mathcal{X}$, $|\mathcal{X}|$ is the total number of elements in $\mathcal{X}$.
$\{0,1\}^n\ (n\in\mathbb{N})$ denotes the set of all $n-$bit strings.
$s_1||s_2$ represents the concatenation of two strings $s_1$ and $s_2$.
Let $\mathcal{W}=\{0,1\}^k\ (k\in\mathbb{N})$ be the keyword space and $\mathcal{ID}=\{0,1\}^{\lambda-1}\setminus\{0^{\lambda-1}\}$ be the file identifier space, where we assume that the string with $(\lambda-1)$-bit zeros, $0^{\lambda-1}$ will never be used as a valid file identifier.
We use the term \emph{entry} to denote the tuple $(op,(w,id))$ of an addition or deletion operation $op\in\{add,del\}$, and a pair of a keyword $w\in\mathcal{W}$ and a file-identifier $id\in\mathcal{ID}$.

\subsection{SEKU Syntax}\label{SEC.SEKU-DEF}

\begin{Definition}[SEKU]\label{DEF.SEKU}
A SEKU scheme $\Sigma$ is composed of four protocols $\textsf{Setup}$, $\textsf{DataUpdate}$, $\textsf{Search}$, and $\textsf{KeyUpdate}$ defined as:
\begin{itemize}
\item $\textsf{Setup}(\lambda)$ takes as input the security parameter $\lambda$, generates the secret key $K_\Sigma$ and private state $\mathbf{State}$ for the client, and initializes the encrypted database $\mathbf{EDB}$ for the server.

\item $\textsf{DataUpdate}(K_\Sigma, \mathbf{State}, op, (w, id); \mathbf{EDB})$ takes as input the secret key $K_\Sigma$ and private state $\mathbf{State}$ to encrypt the entry $(op,(w,id))$ from the client, and finally stores the generated ciphertext into the encrypted database $\mathbf{EDB}$.

\item $\textsf{Search}(K_\Sigma, \mathbf{State}, w; \mathbf{EDB})$ takes as input the secret key $K_\Sigma$, private state $\mathbf{State}$, and a keyword $w$ from the client, and securely delegates the search query of $w$ over the encrypted database $\mathbf{EDB}$ to the server. Finally, this protocol outputs the search results.

\item $\textsf{KeyUpdate}(K_\Sigma, \mathbf{State}; \mathbf{EDB})$ takes as input the secret key $K_\Sigma$ and private state $\mathbf{State}$ from the client, and delegates the key update query to the server. During this protocol, the client updates the secret key $K_\Sigma$ to a new secret key $K_\Sigma^\prime$, and the server executes the key update over $\mathbf{EDB}$ to update the encryption key $K_\Sigma$ of all the stored ciphertexts to $K_\Sigma^\prime$.

\end{itemize}
\end{Definition}

\textbf{Correctness}. We say a SEKU scheme is correct if for any $\lambda\in\mathbb{N}$ and $(K_\Sigma,\mathbf{State};\mathbf{EDB})\leftarrow\textsf{Setup}(\lambda)$, for any $poly(\lambda)$ executions of $\textsf{DataUpdate}(K_\Sigma,\mathbf{State},op,(w,id);\mathbf{EDB})$, $\textsf{Search}(K_\Sigma,\mathbf{State},w;\mathbf{EDB})$, and $\textsf{KeyUpdate}(K_\Sigma, \mathbf{State};$ $ \mathbf{EDB})$, protocol $\textsf{Search}(K_\Sigma,\mathbf{State},w;\mathbf{EDB})$ always returns the set of file identifiers paired with the specific keyword $w$ that have been inserted into $\mathbf{EDB}$ by executing $\textsf{DataUpdate}(K_\Sigma,\mathbf{State},op=add,(w,id);\mathbf{EDB})$ and not yet deleted by executing $\textsf{DataUpdate}(K_\Sigma,\mathbf{State},op=del,(w,id);\mathbf{EDB})$.

Based on Type 1 model, we define the SEKU security against the server and the thief as the indistinguishability of a real and a simulated game. In the real game, the adversary (namely, the server or the thief) runs a real SEKU scheme with adaptively selected inputs, while in the simulated game, the adversary plays with a simulator that simulates a SEKU scheme with a set of leakage functions as inputs. The leakage functions define what information the adversary can infer from observing a real SEKU scheme. If the real game is indistinguishable from the simulated game in the view of the adversary, we say that the leakage of the SEKU scheme to the adversary is strictly bounded by the leakage functions. The following subsections apply the above ideas to formally define the adaptive security of SEKU against the server $\mathcal{A}_\text{Srv}$ and the thief $\mathcal{A}_\text{Thf}$, respectively.

\subsection{Adaptive Security against $\mathcal{A}_\text{Srv}$}

We require that there exists a simulator $\mathcal{S}$ to simulate an ideal game. In the ideal game, server $\mathcal{A}_\text{Srv}$ can adaptively issue queries of $\textsf{Setup}, \textsf{DataUpdate}$, $\textsf{Search}$, and $\textsf{KeyUpdate}$, and simulator $\mathcal{S}$ forges the corresponding responses with leakage functions $\mathcal{L}^{Stp}_\text{Srv},\mathcal{L}^{DaUpdt}_\text{Srv},\mathcal{L}^{Srch}_\text{Srv},$ and $\mathcal{L}^{KeyUpdt}_\text{Srv}$. We say that SEKU is adaptively secure against $\mathcal{A}_\text{Srv}$ if the ideal game is indistinguishable from a real game in the view of $\mathcal{A}_\text{Srv}$. Hence, we have the following formal definition.

\begin{Definition}[Adaptive Security Against $\mathcal{A}_\text{Srv}$]\label{DEF.SS-CKKUA}
Given leakage functions $\mathcal{L}_\text{Srv}=(\mathcal{L}^{Stp}_\text{Srv},\ \mathcal{L}^{DaUpdt}_\text{Srv},\ \mathcal{L}^{Srch}_\text{Srv},$ $\mathcal{L}^{KeyUpdt}_\text{Srv})$, a SEKU scheme $\Sigma$ is said to be $\mathcal{L}_\text{Srv}$-adaptively secure if for any sufficiently large security parameter $\lambda\in\mathbb{N}$ and PPT adversary $\mathcal{A}_\text{Srv}$, there exists an efficient simulator $\mathcal{S}=(\mathcal{S}.\textsf{Setup},\ \mathcal{S}.\textsf{DataUpdate},$ $\mathcal{S}.\textsf{Search},\ \mathcal{S}.\textsf{KeyUpdate})$, such that the probability $|\text{Pr}[Real^\Sigma_{\mathcal{A}_\text{Srv}}(\lambda)=1]-\text{Pr}[Ideal^\Sigma_{\mathcal{A}_\text{Srv}\mathcal{,S},\mathcal{L}_\text{Srv}}(\lambda)=1]|$ is negligible in $\lambda$, where games $Real^\Sigma_{\mathcal{A}_\text{Srv}}(\lambda)$ and $Ideal^\Sigma_{\mathcal{A}_\text{Srv}\mathcal{,S},\mathcal{L}_\text{Srv}}(\lambda)$ are defined as:

\begin{itemize}
\item $Real^\Sigma_{\mathcal{A}_\text{Srv}}(\lambda)$: This real game implements all real SEKU protocols. After initializing $\Sigma$ by running protocol \textsf{Setup}, $\mathcal{A}_\text{Srv}$ adaptively issues \textsf{DataUpdate}, \textsf{Search}, and \textsf{KeyUpdate} queries, and observes the real transcripts of those queries. In the end, $\mathcal{A}_\text{Srv}$ outputs one bit.

\item $Ideal^\Sigma_{\mathcal{A}_\text{Srv}\mathcal{,S},\mathcal{L}_\text{Srv}}(\lambda)$: In this game, $\mathcal{A}_\text{Srv}$ interacts with simulator $\mathcal{S}$ and issues the same queries as in the real game. Simulator $\mathcal{S}$ takes the leakage functions $\mathcal{L}_\text{Srv}=(\mathcal{L}^{Stp}_\text{Srv},\mathcal{L}^{DaUpdt}_\text{Srv},\mathcal{L}^{Srch}_\text{Srv},\mathcal{L}^{KeyUpdt}_\text{Srv})$ as inputs and respectively simulates the corresponding transcripts of SEKU protocols \textsf{Setup}, \textsf{DataUpdate}, \textsf{Search}, and \textsf{KeyUpdate} for $\mathcal{A}_\text{Srv}$ by running $\mathcal{S}.\textsf{Setup}$, $\mathcal{S}.\textsf{DataUpdate},$ $\mathcal{S}.\textsf{Search}$, and $\mathcal{S}.\textsf{KeyUpdate}$. In the end, $\mathcal{A}_\text{Srv}$ outputs one bit.
\end{itemize}
\end{Definition}

\subsection{Adaptive Security against $\mathcal{A}_\text{Thf}$}

This security is also defined as the indistinguishability between a real game and an ideal game. Specifically, we require there exists a simulator $\mathcal{S}^\prime$ to simulate the ideal game. In the simulation process, $\mathcal{S}^\prime$ simulates protocols \textsf{Setup}, \textsf{DataUpdate}, \textsf{Search}, and \textsf{KeyUpdate} with leakage functions $\mathcal{L}^{Stp}_\text{Thf}$, $\mathcal{L}^{DaUpdt}_\text{Thf}$, $\mathcal{L}^{Srch}_\text{Thf}$, and $\mathcal{L}^{KeyUpdt}_\text{Thf}$, respectively.
Besides, $\mathcal{S}^\prime$ forges the key compromise event with leakage function $\mathcal{L}^{KeyLeak}_\text{Thf}$, where $\mathcal{L}^{KeyLeak}_\text{Thf}$ is from a real key-compromise event. We say that SEKU is adaptively secure against $\mathcal{A}_\text{Thf}$ if the ideal game is indistinguishable from a real game in the view of $\mathcal{A}_\text{Thf}$. Formally, we have:

\begin{Definition}[Adaptive Security Against $\mathcal{A}_\text{Thf}$]\label{DEF.SS-KCA}
Given leakage functions $\mathcal{L}_\text{Thf} = ( \mathcal{L}^{Stp}_\text{Thf},\mathcal{L}^{DaUpdt}_\text{Thf},\mathcal{L}^{Srch}_\text{Thf},\mathcal{L}^{KeyUpdt}_\text{Thf},$ $\mathcal{L}^{KeyLeak}_\text{Thf})$, a SEKU scheme $\Sigma$ is $\mathcal{L}_\text{Thf}$-adaptively secure if for any sufficiently large security parameter $\lambda\in\mathbb{N}$ and PPT adversary $\mathcal{A}_\text{Thf}$, there exists an efficient simulator $\mathcal{S}^\prime=(\mathcal{S}^\prime.\textsf{Setup},$ $\mathcal{S}^\prime.\textsf{DataUpdate},\ \mathcal{S}^\prime.\textsf{Search},\ \mathcal{S}^\prime.\textsf{KeyUpdate},\  \mathcal{S}^\prime.\textsf{KeyLeak})$, such that the probability $|\text{Pr}[Real^\Sigma_{\mathcal{A}_\text{Thf}}(\lambda)=1]-\text{Pr}[Ideal^\Sigma_{\mathcal{A}_\text{Thf},\mathcal{S}^\prime,\mathcal{L}_\text{Thf}}(\lambda)=1]|$ is negligible in $\lambda$, where games $Real^\Sigma_{\mathcal{A}_\text{Thf}}(\lambda)$ and $Ideal^\Sigma_{\mathcal{A}_\text{Thf},\mathcal{S}^\prime,\mathcal{L}_\text{Thf}}(\lambda)$ are defined as:

\begin{itemize}
\item $Real^\Sigma_{\mathcal{A}_\text{Thf}}(\lambda)$: This real game exactly implements all SEKU protocols. The thief $\mathcal{A}_\text{Thf}$ adaptively issues \textsf{DataUpdate}, \textsf{Search}, and \textsf{KeyUpdate} queries and observes the real transcripts generated by those queries. In addition, $\mathcal{A}_\text{Thf}$ can adaptively compromise the secret key and private state multiple times. In the end, it outputs one bit.

\item $Ideal^\Sigma_{\mathcal{A}_\text{Thf},\mathcal{S}^\prime,\mathcal{L}_\text{Thf}}(\lambda)$: In this game, the thief $\mathcal{A}_\text{Thf}$ issues the same queries as in the real game.
$\mathcal{S}^\prime$ forges the corresponding transcripts for $\mathcal{A}_\text{Thf}$ by running $\mathcal{S}^\prime.\textsf{Setup}$, $\mathcal{S}^\prime.\textsf{DataUpdate}$, $\mathcal{S}^\prime.\textsf{Search}$, and $\mathcal{S}^\prime.\textsf{KeyUpdate}$ with leakage functions $\mathcal{L}^{Stp}_\text{Thf}$, $\mathcal{L}^{DaUpdt}_\text{Thf}$, $\mathcal{L}^{Srch}_\text{Thf}$, and $\mathcal{L}^{KeyUpdt}_\text{Thf}$, respectively.
When a real key-compromise event happens, $\mathcal{S}^\prime$ runs $\mathcal{S}^\prime.\textsf{KeyLeak}$ with leakage function $\mathcal{L}_\text{Thf}^{KeyLeak}$ to simulate the key-compromise event.
In the end, $\mathcal{A}_\text{Thf}$ outputs one bit.
\end{itemize}
\end{Definition}

\section{Post-Compromise Security of SEKU}\label{SEC.LEAKAGEFUNC}

\subsection{Overview}\label{Section4.1}

As explained in Section \ref{SEC.SECMODEL}, this work focuses on Type 1 security model where the server and the thief do not collude. In this section, we will define the \emph{post-compromise security} of SEKU against Type 1 model by specifying the information leakage allowed to the thief $\mathcal{A}_\text{Thf}$ and the server $\mathcal{A}_\text{Srv}$. To resist the thief $\mathcal{A}_\text{Thf}$, the security must guarantee the privacy of both the ciphertexts and search queries that are generated with non-compromised secret keys, even if any of the historical and future secret keys are compromised. To achieve this goal, we disallow $\mathcal{A}_\text{Thf}$ to learn anything from protocol \textsf{KeyUpdate}. In addition, we also expect that the post-compromise security maintains the confidentiality of the newly generated ciphertexts against $\mathcal{A}_\text{Thf}$, even if those ciphertexts are generated with a compromised key.

When executing protocol \textsf{KeyUpdate}, the server $\mathcal{A}_\text{Srv}$ is delegated to update the secret key of ciphertexts to a new one. It is necessary to guarantee that protocol \textsf{KeyUpdate} does not leak any privacy to $\mathcal{A}_\text{Srv}$.
In other words, the post-compromise security requires that \textsf{KeyUpdate} leaks nothing to $\mathcal{A}_\text{Srv}$, except that $\mathcal{A}_\text{Srv}$ can know if a ciphertext is key-updated from an old (existing) ciphertext.

In summary, our security defines the following basic limitations on information leakage:
\begin{itemize}
\item \textbf{Limit 1}: A compromised key is allowed only to break the ciphertexts that were generated since the last execution of protocol \textsf{KeyUpdate}.
\item \textbf{Limit 2}: Both a \textsf{KeyUpdate} query and its resulting encrypted database leak nothing to the thief $\mathcal{A}_\text{Thf}$.
\item \textbf{Limit 3}: The thief $\mathcal{A}_\text{Thf}$ cannot distinguish any two \textsf{Search} queries.
\item \textbf{Limit 4}: A \textsf{DataUpdate} query leaks nothing to the thief $\mathcal{A}_\text{Thf}$.
Note this implies that a \textsf{DataUpdate} query also leaks nothing to the server $\mathcal{A}_\text{Srv}$.
\item \textbf{Limit 5}: Protocol \textsf{KeyUpdate} leaks (at most) if a ciphertext is key-updated from an old (existing) ciphertext to the server $\mathcal{A}_\text{Srv}$.
\end{itemize}

We note that the security does not restrict the information leakage of \textsf{Search} queries to $\mathcal{A}_\text{Srv}$.
One may design a SEKU instance with the expected information leakage of a \textsf{Search} query to satisfy the security requirements in practice.

\subsection{Definition of Timestamp}

In SEKU, a timestamp denotes when a query is issued. Let $Q^\text{DU}$, $Q^\text{Srch}$, and $Q^\text{KU}$ denote three client query lists: (1) $Q^\text{DU}$ contains all \textsf{DataUpdate} queries in the form of $(u,op,(w,id))$, (2) $Q^\text{Srch}$ records all \textsf{Search} queries in the form of $(u,w)$, and (3) $Q^\text{KU}$ stores all \textsf{KeyUpdate} timestamps in the form of $u$, where $u\in\mathbb{N}$, $op\in\{add,del\}$, $w\in\mathcal{W}$, and $id\in\mathcal{ID}$. We define $U_\text{now}$ as the timestamp of a current query.

Since all timestamps are unique, we can use a timestamp to identify a ciphertext. For example, given the ciphertext $\mathbf{C}$ generated by a \textsf{DataUpdate} query $(u,op,(w,id))\in Q^\text{DU}$, we can use the timestamp $u$ to identify the ciphertext $\mathbf{C}$. Given a $\textsf{KeyUpdate}$ query issued at timestamp $u^\prime$, suppose this query updates $n$ existing ciphertexts $(\mathbf{C}_1,...,\mathbf{C}_n)$ and then inserts the resulted ciphertexts $(\mathbf{C}_1^\prime,...,\mathbf{C}_n^\prime)$ to $\mathbf{EDB}$, we associate ciphertext $\mathbf{C}_i^\prime$ ($i\in[1,n]$) with timestamp $u^\prime+i-1$, and the subsequent queries will start at timestamp $u^\prime+n$.

\subsection{The Formal Definition}\label{SEC.PCS}

We first define several basic leakage functions and then use them to define post-compromise security.

\textbf{Basic Leakage Functions.} Let $u=\textsf{Time}(\mathbf{C})$ denote retrieving the corresponding timestamp $u$ of a given ciphertext $\mathbf{C}$. We use $\textsf{CTRelation}(u^\text{KU})$ to denote the timestamp relationships between the pre-key-updated ciphertexts and the post-key-updated ciphertexts after completing a $\textsf{KeyUpdate}$ query at timestamp $u^\text{KU}\in Q^\text{KU}$. Formally, we have
\begin{align*}
\textsf{CTR}&\textsf{elation}(u^\text{KU})=\\&\{(u,u^\prime)\mid \exists \text{ ciphertexts } \mathbf{C} \text{ and } \mathbf{C}^\prime,~u=\textsf{Time}(\mathbf{C})\text{ and }\\&u^\prime=\textsf{Time}(\mathbf{C}^\prime)\text{ s.t. }\mathbf{C}\text{ is updated from }\mathbf{C}^\prime \text{ by executing }\\&\textsf{KeyUpdate}\text{ at timestamp }u^\text{KU}\}.
\end{align*}

Let $\textsf{KUHist}(u)$ record each \textsf{KeyUpdate} timestamp $u^\text{KU}$ no more than $u$ and the corresponding $\textsf{CTRelation}(u^\text{KU})$, namely
\begin{align*}
\textsf{KUHist}(u)=\{(u^\text{KU},\textsf{CTRel}&\textsf{ation}(u^\text{KU}))\mid \\& u^\text{KU}\in Q^\text{KU}\text{ and }u^\text{KU}\le u\}.
\end{align*}

Let $\textsf{CUHist}(u)$ be the original data of all the ciphertexts generated by executing the \textsf{KeyUpdate} query at the maximum timestamp $u^\text{KU}$ satisfying $u^\text{KU}\le u$. If no such $u^\text{KU}$, $\textsf{CUHist}(u)=\emptyset$. Formally, we have
\begin{align*}
\textsf{CUHi}&\textsf{st}(u)=\\&\{(u^\prime,op,(w,id))\mid \exists\text{ ciphertext }\mathbf{C},~u^\prime=\textsf{Time}(\mathbf{C})\text{ and }\\&(op,(w,id))\text{ is the content of }\mathbf{C}\text{ s.t. }\mathbf{C}\text{ is generated by }\\&\text{executing}~\textsf{KeyUpdate}~\text{at timestamp}~u^\text{KU}~\text{and}~u^\text{KU}\\&\text{satisfying}~u^\text{KU}\le u~\text{is}~\text{the maximum}~\text{one in}~Q^\text{KU}\}.
\end{align*}

Let $\textsf{DUHist}(u)$ denote all the \textsf{DataUpdate} queries issued by the client since the \textsf{KeyUpdate} query of the maximum timestamp $u^\text{KU}$, where $u^\text{KU}$ satisfies $u^\text{KU}\le u$. If no such $u^\text{KU}$, we define $u^\text{KU}=0$. Formally, We have
\begin{align*}
\textsf{DUH}\textsf{ist}(u)=&\{(u^\prime,op,(w,id))\mid (u^\prime,op,(w,id))\in Q^\text{DU}~\text{s.t.}\\&u^\prime>u^\text{KU}~\text{where}~u^\text{KU}\text{satisfying}~u^\text{KU}\le u~\text{is the}\\&\text{maximum one in}~Q^\text{KU}\cup\{0\}\}.
\end{align*}

Next, we define the leakage functions of post-compromise security according to the limitations given in Section~\ref{Section4.1}.

\textbf{Leakage Functions of Key-Compromise}. According to \textbf{Limit 1}, when the secret key is compromised at timestamp $u$, $\mathcal{A}_\text{Thf}$ is only allowed to learn the content of encrypted database $\mathbf{EDB}$ and client queries since the last execution of protocol \textsf{KeyUpdate}. The leakage function $\mathcal{L}^{KeyLeak}_\text{Thf}$ is defined as
\begin{align*}
\mathcal{L}^{KeyLeak}_\text{Thf}=\mathcal{L}^\prime_\text{Thf}(\textsf{CUHist}(U_\text{now}), \textsf{DUHist}(U_\text{now}))
\end{align*}
where $\mathcal{L}^\prime_\text{Thf}$ is a stateless function.

\textbf{Leakage Functions of \textsf{KeyUpdate}.} According to \textbf{Limit 2} and \textbf{Limit 5}, during the \textsf{KeyUpdate} process, the thief cannot obtain any information, and the server can only learn the timestamp relationships between the pre-key-updated ciphertexts and the post-key-updated ciphertexts. We have
\begin{align*}
\mathcal{L}^{KeyUpdt}_\text{Thf}=\text{NULL}~\text{and}~\mathcal{L}^{KeyUpdt}_\text{Srv}=\mathcal{L}_\text{Srv}^\prime(\textsf{KUHist}(U_\text{now}))
\end{align*}
where $\mathcal{L}_\text{Srv}^\prime$ is a stateless function.

\textbf{Leakage Functions of \textsf{Search}.} According to \textbf{Limit 3}, for any two keywords $w_1$ and $w_2$, $\mathcal{L}^{Srch}_\text{Thf}(w_1)$ should be indistinguishable from $\mathcal{L}^{Srch}_\text{Thf}(w_2)$. Namely, the thief cannot obtain any information from the \textsf{Search} process, even if it has compromised the secret key. Thus, the leakage function $\mathcal{L}^{Srch}_\text{Thf}$ is defined as
\begin{align*}
\mathcal{L}^{Srch}_\text{Thf}(w)=\text{NULL}.
\end{align*}

\textbf{Leakage functions of \textsf{DataUpdate}.} According to \textbf{Limit 4}, a newly issued \textsf{DataUpdate} query should leak nothing to both the thief and the server. We define $\mathcal{L}^{DaUpdt}_\text{Thf}$ and $\mathcal{L}^{DaUpdt}_\text{Srv}$ as
\begin{align*}
&\mathcal{L}^{DaUpdt}_\text{Thf}(op,(w,id))=\text{NULL},\\&\mathcal{L}^{DaUpdt}_\text{Srv}(op,(w,id))=\text{NULL}.
\end{align*}

Finally, according to the adaptive security against the thief and the server defined in Section \ref{SEC.DEFINITION} and the above-defined leakage functions, we formalize the post-compromise security as below.
\begin{Definition}[Post-Compromise Security]
A SEKU scheme is post-compromise secure \emph{iff} it is $\mathcal{L}_\text{Srv}$-adaptively secure and $\mathcal{L}_\text{Thf}$-adaptively secure with the following restrictions on the leakage functions $\mathcal{L}_\text{Srv}$ and $\mathcal{L}_\text{Thf}$ simultaneously:
\begin{enumerate}
\item For $\mathcal{L}_\text{Srv}=(\mathcal{L}^{Stp}_\text{Srv},~\mathcal{L}^{DaUpdt}_\text{Srv},~\mathcal{L}^{Srch}_\text{Srv},~\mathcal{L}^{KeyUpdt}_\text{Srv}$),
leakage functions $\mathcal{L}^{DaUpdt}_\text{Srv}$ and $\mathcal{L}^{KeyUpdt}_\text{Srv}$ can be written as:
\begin{gather*}
\mathcal{L}^{DaUpdt}_\text{Srv}(op,(w,id))=\text{NULL},\\
\mathcal{L}^{KeyUpdt}_\text{Srv}=\mathcal{L}^\prime_\text{Srv}(\textsf{KUHist}(U_\text{now})),
\end{gather*}
where $\mathcal{L}_\text{Srv}^\prime$ is a stateless function.
\item For $\mathcal{L}_\text{Thf}\ \ =(\ \ \mathcal{L}^{Stp}_\text{Thf}\ ,\ \ \mathcal{L}^{DaUpdt}_\text{Thf}\ ,\ \ \mathcal{L}^{Srch}_\text{Thf},\ \ \mathcal{L}^{KeyUpdt}_\text{Thf}\ ,$\\ $\mathcal{L}^{KeyLeak}_\text{Thf})$, leakage functions $\mathcal{L}^{DaUpdt}_\text{Thf}$, $\mathcal{L}^{Srch}_\text{Thf}$\ , $\mathcal{L}^{KeyUpdt}_\text{Thf}$, and $\mathcal{L}^{KeyLeak}_\text{Thf}$ can be written as:
\begin{gather*}
\mathcal{L}^{DaUpdt}_\text{Thf}(op,(w,id))=\text{NULL},\\
\mathcal{L}^{Srch}_\text{Thf}(w)=\text{NULL},~
\mathcal{L}^{KeyUpdt}_\text{Thf}=\text{NULL},\\
\mathcal{L}^{KeyLeak}_\text{Thf}=\mathcal{L}^\prime_\text{Thf}(\textsf{CUHist}(U_\text{now}),\textsf{DUHist}(U_\text{now})),
\end{gather*}
where $\mathcal{L}^\prime_\text{Thf}$ is a stateless function.
\end{enumerate}
Note that the post-compromise security does not put any restriction on leakage functions $\mathcal{L}^{Stp}_\text{Srv}$, $\mathcal{L}^{Srch}_\text{Srv}$, and $\mathcal{L}^{Stp}_\text{Thf}$.
\end{Definition}

The definition of leakage function $\mathcal{L}^{DaUpdt}_\text{Srv}$ in the post-compromise security also satisfies the notion of forward security in the context of DSSE~\cite{DBLP:conf/ndss/StefanovPS14, DBLP:conf/ccs/Bost16}.
Specifically, the forward security requires that \emph{a \textsf{DataUpdate} query does not leak any information about the updated keywords}.
The post-compromise security defines the leakage functions of \textsf{DataUpdate} for both $\mathcal{A}_\text{Srv}$ and $\mathcal{A}_\text{Thf}$ are NULL.
Such a definition clearly satisfies the forward security. Hence, we state that post-compromise security subsumes forward security.
It is worth mentioning that the post-compromise security does not explicitly limit the leakage function $\mathcal{L}^{Srch}_\text{Srv}$.
One may design a SEKU instance with different $\mathcal{L}^{Srch}_\text{Srv}$ to achieve various strengths of security against the server, e.g., backward security. Thus, post-compromise security is compatible with backward security.

\section{\texttt{Bamboo}: A SEKU Instance}\label{SEC.BAMBOO}

This section presents \texttt{Bamboo}, a post-compromise-secure SEKU instance that achieves constant \textsf{DataUpdate} complexity, sub-linear \textsf{Search} overhead, and non-interactive \textsf{KeyUpdate}. \texttt{Bamboo} leverages the two-layer encryption and inter-ciphertext chain-like structure techniques to generate its ciphertexts.

\subsection{Building Blocks}

The construction of \texttt{Bamboo} relies on an invertible mapping function and the Diffie-Hellman key exchange protocol. This part briefly introduces these building blocks.

\textbf{Invertible Mapping Function.} The invertible mapping function $\pi:\{0,1\}^n\rightarrow\mathbb{G}$ probabilistically maps an $n-$bit string to an element of a multiplicative cyclic group $\mathbb{G}$ of prime order $q$, and $\pi^{-1}$ denotes the deterministic inverse of $\pi$. We use $(n,\mathbb{G},q,\pi,\pi^{-1})\leftarrow\mathbf{PGen}(\lambda,n)$ to denote an efficient probabilistic algorithm that takes as input a security parameter $\lambda$ and bit length $n$ of strings, and initializes the invertible mapping function $\pi$ and its inverse $\pi^{-1}$. We assume that the DDH assumption~\cite{DBLP:conf/ants/Boneh98} (defined in Appendix~\ref{APP.DDH}) holds in the group $\mathbb{G}$. Boyd et al.~\cite{DBLP:conf/crypto/BoydDGJ20} explained how to implement such probabilistic $\pi$ by embedding a bit string into the X-coordinate of an elliptic curve point.

\textbf{Diffie-Hellman Key Exchange Protocol~\cite{DBLP:journals/tit/DiffieH76}.} This protocol enables the client and the server to generate a shared key via an insecure communication channel without prior shared secrets. 
We briefly introduce the elliptic curve variant of the Diffie-Hellman key exchange protocol.
The client and the server first initialize an elliptic curve group $\mathbb{G}^\prime$ whose order is prime $q^\prime$. Let $g^\prime$ be a generator of $\mathbb{G}^\prime$. Then, the client samples a secret random number $a\in\mathbb{Z}_{q^\prime}^*$ and the server chooses a secret random number $b\in\mathbb{Z}_{q^\prime}^*$. Next, the client and the server compute and exchange ${g^\prime}^a$ and ${g^\prime}^b$, respectively. Finally, they compute the shared key $({g^\prime}^b)^a=({g^\prime}^a)^b$.
After that, they can use the shared key to establish a secure channel.
In the construction of \texttt{Bamboo}, when running the key change protocol, we assume the secret random numbers and the shared key are ephemeral. Namely, they will be permanently deleted after the secure channel is closed and will never be leaked out the memories of the client and the server.

\begin{algorithm}[h]
\caption{Protocols \textsf{Setup} and \textsf{DataUpdate} of \texttt{Bamboo}.}\label{ALG.SEKU-1}
\leftline{\underline{$\textsf{Setup}(\lambda, a_\text{max})$}}
\begin{algorithmic}[1]
\STATE Initialize an invertible mapping function and its inverse $(\lambda,\mathbb{G},q,\pi,$ $\pi^{-1})\leftarrow\mathbf{PGen}(\lambda,\lambda)$
\STATE Initialize three cryptographic hash functions $\mathbf{H}_1:\{0,1\}^*\rightarrow \mathbb{G}$, $\mathbf{H}_2:\{0,1\}^*\rightarrow\mathbb{G}$, and $\mathbf{G}:\{0,1\}^*\rightarrow \mathbb{G}$
\STATE Initialize the Diffie-Hellman key exchange protocol parameters including the elliptic curve group $\mathbb{G}^\prime$ of prime order $q^\prime$ and a generator $g^\prime\in\mathbb{G}^\prime$
\STATE Initialize two empty maps $\mathbf{State}\leftarrow\emptyset$ and $\mathbf{EDB}\leftarrow\emptyset$
\STATE Initialize the secret key $K_\Sigma=(K_1,K_2)\overset{\$}{\leftarrow} \mathbb{Z}^*_q\times \mathbb{Z}^*_q$
\STATE Send the encrypted database $\mathbf{EDB}$ to the server
\end{algorithmic}
\leftline{\underline{$\textsf{DataUpdate}(K_\Sigma, \mathbf{State}, op, (w, id);\mathbf{EDB})$}}
\leftline{Client:}
\begin{algorithmic}[1]
\STATE Retrieve record $(tk_w,cnt_w)$ from $\mathbf{State}[w]$
\IF{$(tk_w,cnt_w)=(\text{NULL},\text{NULL})$}
\STATE Randomly draw $tk_w\leftarrow \{0,1\}^\lambda$ and initialize $cnt_w\leftarrow 0$
\ENDIF
\STATE Accumulate $cnt_w\leftarrow cnt_w+1$
\STATE Randomly sample $tk^\prime_w\leftarrow \{0,1\}^\lambda$
\STATE Compute ciphertext label $L\leftarrow (\mathbf{H}_1(tk^\prime_w))^{K_1}$
\STATE Encrypt search token $D\leftarrow (\pi(tk_w)\cdot \mathbf{H}_2(tk^\prime_w))^{K_1}$
\STATE Compute component $C\leftarrow (\pi(op||id))^{K_2}\cdot (\mathbf{G}(tk^\prime_w))^{K_1}$
\STATE Update private state $\mathbf{State}[w]\leftarrow (tk^\prime_w, cnt_w)$
\STATE Send the ciphertext $(L,D,C)$ to the server\\
\vspace{2pt}
\leftline{Server:}
\STATE Store $\mathbf{EDB}[L]\leftarrow (D,C)$
\end{algorithmic}
\end{algorithm}

\subsection{The Construction}

Algorithms~\ref{ALG.SEKU-1},~\ref{ALG.SEKU-2}, and~\ref{ALG.SEKU-3} present the details of \texttt{Bamboo}.
To initialize the scheme, protocol \textsf{Setup} takes as inputs a security parameter $\lambda$ and a maximum padding value $a_\text{max}$. Protocol \texttt{Bamboo}.\textsf{Setup} initializes an invertible mapping function, three cryptographic hash functions, and the Diffie-Hellman key exchange protocol parameters.
The above-initialized parameters and functions can be publicly known.
Then the client initializes a search key $K_1$, an encryption key $K_2$, an empty private state $\mathbf{State}$, and an empty database $\mathbf{EDB}$.
Finally, it sends $\mathbf{EDB}$ to the server.

{\bf \textsf{DataUpdate}}. To encrypt an entry $(op,(w,id))$, the client executes protocol \textsf{DataUpdate}. It generates a ciphertext $(L,D,C)$ and sends the generated ciphertext to the server. When generating $L$, $D$, and $C$, the cryptographic hash functions implement the first layer encryption, and the exponentiation operations over group $\mathbb{G}$ implement the second layer encryption. Label $L$ indexes this ciphertext in $\mathbf{EDB}$. Component $D$ encrypts a search token $tk_w$ of the prior ciphertext. The server can use $tk_w$ to find the prior ciphertext and decrypt the search token encrypted in it. Component $C$ encrypts $op||id$.

\begin{algorithm}[h]
\caption{Protocol \texttt{Bamboo}.\textsf{Search}.}\label{ALG.SEKU-2}
\leftline{\underline{$\textsf{Search}(K_\Sigma, \mathbf{State}, w;\mathbf{EDB})$}}
\leftline{Client:}
\begin{algorithmic}[1]
\STATE Retrieve record $(tk_w,cnt_w)$ from $\mathbf{State}[w]$
\STATE Abort if both $tk_w$ and $cnt_w$ are NULL
\STATE Establish a temporary secure channel with the server using the Diffie-Hellman key exchange protocol
\STATE Compute ciphertext label $L\leftarrow(\mathbf{H}_1(tk_w))^{K_1}$
\STATE Compute $Msk_D\leftarrow (\mathbf{H}_2(tk_w))^{K_1}$ and $Msk_C\leftarrow (\mathbf{G}(tk_w))^{K_1}$
\STATE Send search trapdoor $(K_1,L,Msk_D,Msk_C)$ to the server via above secure channel\\
\vspace{2pt}
\leftline{Server:}
\STATE Initialize an empty list $\mathcal{I}\leftarrow\emptyset$
\STATE Retrieve $(D,C)\leftarrow\mathbf{EDB}[L]$
\WHILE{$(D,C)\neq(\text{NULL},\text{NULL})$}
\STATE Decrypt search token $tk\leftarrow \pi^{-1}((\frac{D}{Msk_D})^{K_1^{-1}})$
\STATE Compute $C^\prime\leftarrow\frac{C}{Msk_C}$ and insert $C^\prime$ into $\mathcal{I}$
\STATE Compute $L\leftarrow (\mathbf{H}_1(tk))^{K_1}$, $Msk_D\leftarrow(\mathbf{H}_2(tk))^{K_1}$, and $Msk_C\leftarrow(\mathbf{G}(tk))^{K_1}$
\STATE Retrieve $(D,C)\leftarrow\mathbf{EDB}[L]$
\ENDWHILE
\STATE Let $n$ be the number of found ciphertexts
\STATE Pad $\mathcal{I}$ with $a_\text{max}-n$ arbitrary elements of $\mathbb{G}$
\STATE Return $\mathcal{I}$ to the client via the above secure channel\\
\vspace{2pt}
\leftline{Client:}
\STATE Initialize an empty list $\mathcal{R}$
\FOR{$i=1$ \TO $cnt_w$}
\STATE Decrypt the $i$-th component $C^\prime_i$ of $\mathcal{I}$ by computing $op_i||id_i\leftarrow \pi^{-1}({C^\prime}_i^{K_2^{-1}})$
\STATE If $op_i=add$, insert $id_i$ into $\mathcal{R}$. Otherwise delete $id_i$ from $\mathcal{R}$.
\ENDFOR
\STATE \textbf{return} $\mathcal{R}$
\end{algorithmic}
\end{algorithm}

{\bf \textsf{Search}}. To perform the secure search for a keyword $w$, the client executes \textsf{Search}. The client first retrieves the search token $tk_w$ and the \textsf{DataUpdate} counter $cnt_w$ of keyword $w$ from $\mathbf{State}$. If both $tk_w$ and $cnt_w$ are NULL, namely, the client has never issued a \textsf{DataUpdate} query about $w$, the \textsf{Search} process aborts (Steps 1 and 2). With the search token $tk_w$, the client computes the label $L$ of the latest generated ciphertext $\mathbf{C}=(L,D,C)$ containing $w$. Then the client computes $Msk_D$ and $Msk_C$ with $tk_w$. The server can use $Msk_D$ and $Msk_C$ to decrypt the search token from the component $D$ and partially decrypt the component $C$ of the latest ciphertext $(L,D,C)$ (Steps 4 and 5). Finally, the client sends the search trapdoor $(K_1,L,Msk_D,Msk_C)$ to the server. Upon receiving the search trapdoor, the server uses label $L$ to retrieve the latest issued ciphertext $(L,D,C)$ and uses the search key $K_1$, $Msk_D$, and $Msk_C$ to decrypt the search token $tk$ of the prior ciphertext and partially decrypt the component $C$, respectively. Then, the server uses the decrypted search token $tk$ to locate and decrypt the prior ciphertext. In this way, the server traverses the hidden chain from the latest issued ciphertext of $w$ and finds all the matching ciphertexts (Steps 8 to 14).
Then the server pads the size of search results to $a_\text{max}$ and returns them to the client.
Finally, the client decrypts the first $cnt_w$ ciphertexts with encryption key $K_2$ and filters out invalid file identifiers according to their operations $op$.

\begin{algorithm}[t]
\caption{Protocol \texttt{Bamboo}.\textsf{KeyUpdate}.}\label{ALG.SEKU-3}
\leftline{\underline{$\textsf{KeyUpdate}(K_\Sigma,\mathbf{State};\mathbf{EDB})$}}
\leftline{Client:}
\begin{algorithmic}[1]
\STATE Establish a temporary secure channel with the server using the Diffie-Hellman key exchange protocol
\STATE Randomly draw the \textsf{KeyUpdate} token $\Delta$ from $\mathbb{Z}^*_q$
\STATE Update local secret keys $K_1\leftarrow K_1\cdot\Delta$ and $K_2\leftarrow K_2\cdot\Delta$
\STATE Send $\Delta$ to the server via the above secure channel\\
\vspace{2pt}
\leftline{Server:}
\FOR{All $(L,D,C)\textbf{ such that }(D,C)\leftarrow\mathbf{EDB}[L]$}
\STATE Update label $L^\prime\leftarrow L^{\Delta}$
\STATE Update encrypted search token $D^\prime\leftarrow D^\Delta$
\STATE Update component $C^\prime\leftarrow C^{\Delta}$
\STATE Insert ciphertext $\mathbf{EDB}[L^\prime]\leftarrow (D^\prime,C^\prime)$
\STATE Remove ciphertext $(L,D,C)$ from $\mathbf{EDB}$
\ENDFOR
\end{algorithmic}
\end{algorithm}

{\bf \textsf{KeyUpdate}}. To update the secret key, the client executes \textsf{KeyUpdate}. The client first samples a random element $\Delta$ from $\mathbb{Z}^*_q$, and then updates search key $K_1$ and encryption key $K_2$ by multiplying those keys by $\Delta$. Then, the client sends the \textsf{KeyUpdate} token $\Delta$ to the server. Finally, the server updates the key of the whole encrypted database using $\Delta$.

{\bf Complexity \& Cost.} \texttt{Bamboo} achieves constant \textsf{DataUpdate} time cost, sub-linear \textsf{Search} complexity, and linear \textsf{KeyUpdate} complexity.
The computational complexity of \textsf{DataUpdate}, \textsf{Search}, and \textsf{KeyUpdate} are $O(1)$, $O(a_w)$, and $O(N)$, respectively, where symbol $a_w$ is the total number of \textsf{DataUpdate} queries of searched keyword $w$, and $N$ is the size of the encrypted database $\mathbf{EDB}$.
In terms of bandwidth cost, protocols \textsf{DataUpdate}, \textsf{Search}, and \textsf{KeyUpdate} exchange $O(1)$, $O(a_\text{max})$, and $O(1)$ data between the client and the server, respectively.
We will evaluate \texttt{Bamboo}'s practical efficiency in Section~\ref{SEC.EXPERIMENT}. More discussions on \texttt{Bamboo} can be found in Section \ref{Diss.BambooCons}.

\begin{figure*}
\centering
\includegraphics[width=0.98\linewidth]{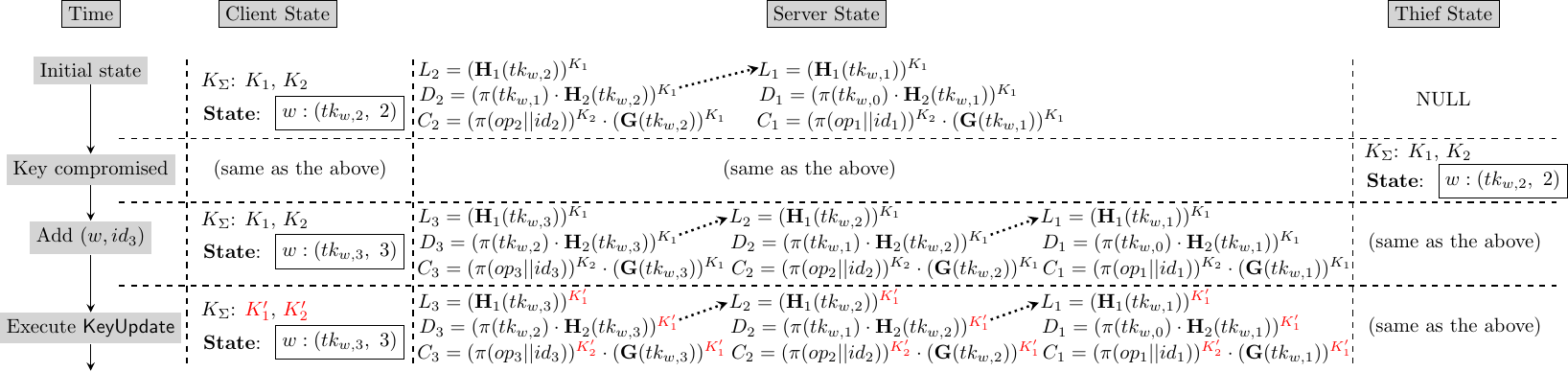}
\caption{An example of \texttt{Bamboo}. In the beginning, the client has run \textsf{DataUpdate} with $(op_1,(w,id_1))$ and $(op_2,(w,id_2))$. 
After the compromise, if the client is not ``warned" immediately, it may still perform a new \textsf{DataUpdate} query on $(op_3,(w,id_3))$. At last, the client executes \textsf{KeyUpdate} for $K_\Sigma$.}\label{F.BambooExample}
\end{figure*}

\subsection{An Example of \texttt{Bamboo}}

We give a concrete example for \texttt{Bamboo} in Figure \ref{F.BambooExample}.
In the beginning, the client holds $K_\Sigma=(K_1, K_2)$ and the private state $(tk_{w,2},2)$ of keyword $w$, where $tk_{w,2}$ is the search token of $w$'s latest generated ciphertext and $2$ is the \textsf{DataUpdate} counter value of $w$.
The server stores $\mathbf{C}_1=(L_1,D_1,C_1)$ and $\mathbf{C}_2=(L_2,D_2,C_2)$ under $w$, in which $\mathbf{C}_1$ encrypts $op_1||id_1$, and $\mathbf{C}_2$ encrypts $op_2||id_2$ and encapsulates the search token $tk_{w,1}$ of  $\mathbf{C}_1$.
Next, the thief compromises $K_\Sigma=(K_1,K_2)$ and the private state ``$w:(tk_{w,2},\ 2)$''.
It can accordingly extract the information $(op_1,(w,id_1))$ and $(op_2,(w,id_2))$ from $\mathbf{C}_1$ and $\mathbf{C}_2$, respectively.

Suppose the client does not receive any warnings about the compromise. It continues to run \textsf{DataUpdate} to encrypt $(op_3, (w,id_3))$ with a randomly selected search token $tk_{w,3}$ and $K_\Sigma$.
The thief knows neither the current information of the private state nor $tk_{w,3}$.
Thus, it does not know to which keyword the ciphertext corresponds and cannot extract the information from the newly generated $\mathbf{C}_3=(L_3,D_3,C_3)$.
This ciphertext leaks nothing to the thief.
At last, the client updates its key $K_\Sigma$ from $(K_1,K_2)$ to $(K_1^\prime, K_2^\prime)$.
The thief cannot use $(K_1,K_2)$ to decrypt any ciphertexts under $(K_1^\prime, K_2^\prime)$.

After the \textsf{KeyUpdate}, the client can use $K_1^\prime$ and $tk_{w,3}$ to generate the trapdoor for a new search query.
The server first locates and decrypts $\mathbf{C}_3$ to obtain $tk_{w,2}$.
It further uses $K^\prime_1$ and $tk_{w,2}$ to identify and decrypt $\mathbf{C}_2$ to get $tk_{w,1}$.
Similarly, the server traverses back to the first ciphertext $\mathbf{C}_1$ (along the chain).
It eventually returns the partially decrypted components $C_1$, $C_2$, and $C_3$ to the client who will fully recover  $(op_1,(w,id_1))$, $(op_2, (w,id_2))$, and $(op_3,(w,id_3))$.

\subsection{Correctness and Security Analysis}

\textbf{Correctness.} The correctness of \texttt{Bamboo} comes from the collision-resistance of hash functions $\mathbf{H}_1$, $\mathbf{H}_2$, and $\mathbf{G}$, the algebraic features of group $\mathbb{G}$, and the correctness of the invertible mapping function $\pi$.
Specifically, when executing $\textsf{DataUpdate}$ with a given entry $(op,(w,id))$, both the collision-resistance of $\mathbf{G}$ and $\mathbf{H}_2$ and the correctness of $\pi$ guarantee that the generated ciphertext $(L,D,C)$ encrypts $op$ and $id$ in component $C$ and encrypts the previously issued ciphertext's search token $tk_w$ in component $D$ correctly.

When executing \textsf{Search} with a given keyword $w$, the latest issued ciphertext $(L,D,C)$ can be correctly located by the label $L=(\mathbf{H}_1(tk_w))^{K_1}$ in the search trapdoor.
Given $K_1$, $Msk_D$, and $Msk_C$ contained in the search trapdoor, the server can decrypt the search token $tk$ of the prior ciphertext from $D$ and partially decrypt component $C$ to get $(\pi(op||id))^{K_2}$.
Then, the server can use search token $tk$ to locate and decrypt the prior ciphertext with hash functions $\mathbf{H}_1$, $\mathbf{H}_2$, and $\mathbf{G}$.
In this way, the server can precisely find all matching ciphertexts and return the correctly partially decrypted ciphertexts to the client. Finally, the client can use the encryption key $K_2$ to decrypt returned results.

Without loss of generality, suppose there exists a ciphertext $(L,D,C)=(L_0^{K_1},D_0^{K_1},C_0^{K_2}\cdot X_0^{K_1})$, where $K_1$ is the search key, and $K_2$ is the encryption key.
After executing protocol \textsf{KeyUpdate} with a \textsf{KeyUpdate} token $\Delta$, the post-key-updated ciphertext is $(L_0^{K_1\cdot\Delta},D_0^{K_1\cdot\Delta},$ $C_0^{K_2\cdot\Delta}\cdot X_0^{K_1\cdot\Delta})$, and the two new keys are $K_1\cdot\Delta$ and $K_2\cdot\Delta$. Let $K_1^\prime=K_1\cdot\Delta$ and $K_2^\prime=K_2\cdot\Delta$.
It is clear that the new keys $K_1^\prime$ and $K_2^\prime$ can be used to search and decrypt the post-key-updated ciphertext.

\textbf{Security.} The post-compromise security of \texttt{Bamboo} against Type 1 model is captured according to the views of the server $\mathcal{A}_\text{Srv}$ and the thief $\mathcal{A}_\text{Thf}$.
For $\mathcal{A}_\text{Srv}$, protocol \textsf{Setup} leaks only the security parameter $\lambda$ and the maximum padding value $a_\text{max}$; protocol \textsf{DataUpdate} leaks nothing; when issuing a search query of keyword $w$, protocol \textsf{Search} leaks search pattern $\textsf{sp}(w)$, file identifiers matching $w$ and the insertion timestamps of those file identifiers (i.e., $\textsf{TimeDB}(w)$), the \textsf{DataUpdate} timestamps of $w$ (i.e., $\textsf{DUTime}(w)$), and the \textsf{KeyUpdate} histories $\textsf{KUHist}(U_\text{now})$; protocol \textsf{KeyUpdate} leaks nothing but $\textsf{KUHist}(U_\text{now})$.
The formal definitions of leakage functions $\textsf{sp}(w)$, $\textsf{TimeDB}(w)$, and $\textsf{DUTime}(w)$ are described as:
\begin{gather*}
\textsf{TimeDB}(w)=\{(u,id)\mid (u,add,(w,id))\in Q^\text{DU}\text{ and }\\
\qquad\qquad\qquad\qquad\forall u^\prime,(u^\prime,del,(w,id))\notin Q^\text{DU} \},\\
\textsf{sp}(w)=\{u\mid (u,w)\in Q^\text{Srch}\},\\
\textsf{DUTime}(w)=\{u\mid (u,op,(w,id))\in Q^\text{DU}\}.
\end{gather*}

As for $\mathcal{A}_\text{Thf}$, protocol \textsf{Setup} leaks the security parameter $\lambda$ and the maximum padding value $a_\text{max}$; protocols \textsf{DataUpdate}, \textsf{Search}, and \textsf{KeyUpdate} leak nothing; when the secret key is compromised, the thief $\mathcal{A}_\text{Thf}$ learns the plaintexts of the ciphertexts that were generated by the last execution of protocol \textsf{KeyUpdate} and the \textsf{DataUpdate} queries issued since the last execution of \textsf{KeyUpdate}.

Formally, we have the following Theorem~\ref{THEO.SEKU-SEC}, whose proof is in Appendix \ref{SEC.SEKU-PROOF}.
Moreover, the leakage functions of \texttt{Bamboo} in the view of the server satisfy the forward security and the backward security in the sense that protocol \textsf{Search} only leaks \emph{the file-identifiers currently matching the queried keyword $w$, when they were uploaded, and when all the \textsf{DataUpdate} on $w$ happened}~\cite{DBLP:conf/ccs/BostMO17} to the server.

\begin{Theorem}\label{THEO.SEKU-SEC}
Suppose hash functions $\mathbf{H}_1$, $\mathbf{H}_2$, and $\mathbf{G}$ are random oracles, and DDH assumption holds in $\mathbb{G}$, $\texttt{Bamboo}$ is a post-compromise-secure SEKU scheme since:
\begin{enumerate}
\item \texttt{Bamboo} is $\mathcal{L}_\text{Srv}$-adaptively secure and the leakage functions $\mathcal{L}_\text{Srv}=(\mathcal{L}^{Stp}_\text{Srv},\mathcal{L}^{DaUpdt}_\text{Srv},$ $\mathcal{L}^{Srch}_\text{Srv},\mathcal{L}^{KeyUpdt}_\text{Srv})$ can be written as:
\begin{gather*}
\mathcal{L}^{Stp}_\text{Srv}(\lambda,a_\text{max})=(\lambda,a_\text{max}),\\
\mathcal{L}^{DaUpdt}_\text{Srv}(op,(w,id))=\text{NULL},\\
\mathcal{L}^{KeyUpdt}_\text{Srv}=\mathcal{L}^\prime_\text{Srv}(\textsf{KUHist}(U_\text{now})),\\
\mathcal{L}^{Srch}_\text{Srv}(w)=\mathcal{L}^{\prime\prime}_\text{Srv}(\textsf{sp}(w),\textsf{TimeDB}(w),\qquad\qquad\qquad\qquad\\
\qquad\qquad\qquad\qquad\textsf{DUTime}(w),\textsf{KUHist}(U_\text{now})),
\end{gather*}
where $\mathcal{L}_\text{Srv}^\prime$ and $\mathcal{L}_\text{Srv}^{\prime\prime}$ are two stateless functions.
\item \texttt{Bamboo} is $\mathcal{L}_\text{Thf}$-adaptively secure and the leakage functions $\mathcal{L}_\text{Thf}=(\mathcal{L}^{Stp}_\text{Thf},\mathcal{L}^{DaUpdt}_\text{Thf},\mathcal{L}^{Srch}_\text{Thf},$ $\mathcal{L}^{KeyUpdt}_\text{Thf},\mathcal{L}^{KeyLeak}_\text{Thf})$ can be written as:
\begin{gather*}
\mathcal{L}^{Stp}_\text{Thf}(\lambda,a_\text{max})=(\lambda,a_\text{max}),\\
\mathcal{L}^{DaUpdt}_\text{Thf}(op,(w,id))=\text{NULL},\\
\mathcal{L}^{Srch}_\text{Thf}(w)=\text{NULL},~
\mathcal{L}^{KeyUpdt}_\text{Thf}=\text{NULL},\\
\mathcal{L}^{KeyLeak}_\text{Thf}=\mathcal{L}^\prime_\text{Thf}(\textsf{CUHist}(U_\text{now}),\textsf{DUHist}(U_\text{now})),
\end{gather*}
where $\mathcal{L}^\prime_\text{Thf}$ is a stateless function.
\end{enumerate}
\end{Theorem}

\subsection{Improvements on Bandwidth}\label{SEC.IMPROVBAND}

In Section~\ref{SEC.PCS}, we define leakage functions $\mathcal{L}^{Srch}_\text{Thf}$ and $\mathcal{L}^{KeyLeak}_\text{Thf}$ for the thief to be stateless.
To achieve this stateless property, a post-compromise-secure SEKU instance has to strictly protect the result volume of each search query from $\mathcal{A}_\text{Thf}$.
Thus, \texttt{Bamboo} pads the size of search results to a pre-defined maximum padding value $a_\text{max}$.
Moreover, this padding clearly causes extra bandwidth.

To reduce this cost while protecting search queries, we propose to apply a flexible padding technique to protocol \textsf{Search}.
We consider the adjustable padding technique introduced by Demertzis et al.~\cite{DBLP:conf/uss/DemertzisPPS20}.

\textbf{Adjustable Padding Overview.} Demertzis et al.~\cite{DBLP:conf/uss/DemertzisPPS20} applied this technique to build a secure static searchable encryption scheme.
Specifically, given a parameter $x (x\ge 2)$ and a static database $\mathbf{DB}$, let $\mathbf{DB}(w)$ be the set of corresponding file identifiers to keyword $w$.
When encrypting $\mathbf{DB}(w)$ for a keyword $w$ in the \textsf{Setup} process, the client finds an integer $i$ such that $x^{i-1}<|\mathbf{DB}(w)|\le x^i$ and pad $x^i-|\mathbf{DB}(w)|$ dummy entries to the encrypted results of $\mathbf{DB}(w)$.
The adjustable padding technique guarantees that the search on a keyword $w$ only leaks to the server the result volume of size $\text{log}_x|\mathbf{DB}(w)|+1$, namely, leaking only $\text{log}_2\text{log}_x|\mathbf{DB}(w)|+1$ bits information.

Unfortunately, we cannot apply the above adjustable padding technique directly to \texttt{Bamboo} since it may lead to a severe security problem.
For example, suppose $\mathcal{A}_\text{Thf}$ compromises the secret key, and there is a keyword $w^\prime$ with a unique adjustable padding value $pad_\text{uni}$.
In this case, when observing that the search results of a client's \textsf{Search} query has the size of $pad_\text{uni}$, the thief $\mathcal{A}_\text{Thf}$ may have a high probability of determining that the client is searching for $w^\prime$.

\begin{algorithm}[h]
\algsetup{linenosize=\small} \small
\caption{Function $\textsf{PaddingVal-adj}(a_\text{max},x, w,\mathbf{State})$.}\label{FUNC.ADJPADING}
\begin{algorithmic}[1]
\STATE Retrieve $(tk_w,cnt_w)\leftarrow \mathbf{State}[w]$
\STATE Find an integer $i$ such that $x^{i-1}< cnt_w \le x^i$
\STATE If there is not a second keyword $w^\prime$ of which \textsf{DataUpdate} counter $cnt_{w^\prime}$ satisfies $x^{i-1}<cnt_{w^\prime}\le x^i$, return $a_\text{max}$
\STATE If $x^i>a_\text{max}$, return $a_\text{max}$
\STATE Randomly return $a_\text{max}$ or $x^i$
\end{algorithmic}
\end{algorithm}

To tackle this problem, we propose a new padding method named \textsf{PaddingVal-adj} (see Algorithm~\ref{FUNC.ADJPADING}).
It takes the maximum padding value $a_\text{max}$, an integer $x$ $(x\ge 2)$, a keyword $w$, and private state $\mathbf{State}$ as inputs, and calculates the adjustable padding value $x^i$ for keyword $w$ according to \textsf{DataUpdate} counter $cnt_w$.
Next, it checks if $x^i$ can be used to \emph{deterministically} distinguish $w$ from other keywords and returns $a_\text{max}$ if so.
Otherwise, the function returns $a_\text{max}$ if $x^i>a_\text{max}$.
Finally, the function randomly chooses and returns one of $x^i$ and $a_\text{max}$ as the padding value.
\textsf{PaddingVal-adj} hides more information than the adjustable padding technique from the thief $\mathcal{A}_\text{Thf}$ who may have (prior) plaintext knowledge about the encrypted database.
Besides, by applying the function, a \textsf{Search} query can take less bandwidth to be completed than using the maximum padding technique.

Note that Step 3 (in Algorithm 4) yields the \textsf{Search} leakage to the server.
Specifically, this step determines if the function returns the maximum padding value $a_\text{max}$ based on the keyword frequency of $w$.
On the other hand, the frequency could possibly be leaked to the server through the padding value.
The ``random return" strategy in Step 5 (Algorithm 4) is to reduce the above leakage by weakening the server's ability to check whether $a_\text{max}$ is selected randomly or computed from the keyword frequency.

For example, suppose the client's encrypted database only contains $w_1$ and $w_2$, and the (search) response sizes for both keywords are $x^{i^\prime}$ for certain $x$ and $i^\prime$.
After the client issues \textsf{Search} queries on $w_1$ and $w_2$, the server can extract from the \textsf{Search} leakage that: (1) the client only uses two distinct keywords in the database and (2) the keywords have the same response size of $x^{i^\prime}$.
Note the server here cannot directly see $w_1$ and $w_2$.
In this context, we explain what will happen if Step 5 deterministically returns the adjustable value.
Upon searching for $w_1$ and $w_2$, the client should require the server to pad the responses to $x^{i^\prime}$.
If the client issues a \textsf{DataUpdate} query and then a \textsf{Search} query on $w_1$ (or $w_2$), the server can easily learn the information about the \textsf{DataUpdate} query. 
More concretely, if the \textsf{Search} query requires the server to pad the results to $x^{i^\prime}$, the server will know the  \textsf{DataUpdate} query contains a keyword different from $w_1$ and $w_2$.
This is because the condition of Step 3 does not hold.
Otherwise (i.e., if the requested padding size is $a_\text{max}$), the server learns that two queries contain distinct keywords, for example, the \textsf{DataUpdate} query contains $w_1$ while $w_2$ is in the \textsf{Search} query, or the other way round.
We note that the ``random" strategy helps us to hide the above connections between keywords and queries from the server.

One can easily apply function \textsf{PaddingVal-adj} in protocol \textsf{Search} of \texttt{Bamboo} to reduce the bandwidth. Specifically, in protocol \textsf{Setup}, the client takes an integer $x(x\ge 2)$ as an additional input.
In protocol \textsf{Search}, before sending a search trapdoor to the server, the client executes function \textsf{PaddingVal-adj} with the maximum padding value $a_\text{max}$, integer $x$, the queried keyword $w$, and the private state $\mathbf{State}$ as inputs to compute the padding value $num_\text{pad}$.
Then, the client sends $num_\text{pad}$ along with the search trapdoor to the server. After finding all matching ciphertexts, the server pads the size of the search results to $num_\text{pad}$ and returns the padded search results to the client. For convenience, we name the resulting scheme $\texttt{Bamboo}^*$.
Section~\ref{SEC.EXPERIMENT} will experimentally test and compare the \textsf{Search} performance of $\texttt{Bamboo}^*$ and $\texttt{Bamboo}$.

Compared with \texttt{Bamboo}, $\texttt{Bamboo}^*$ leaks more information to both $\mathcal{A}_\text{Srv}$ and $\mathcal{A}_\text{Thf}$. For $\mathcal{A}_\text{Srv}$, the padding value during a search should be included in the \textsf{Search} leakage function.
In practice, $\texttt{Bamboo}^*$ is feasible to handle large-scale databases, such as Wikipedia.
When doing so, it is hard for $\mathcal{A}_\text{Srv}$ to infer information from unsearched ciphertexts with padding values.
In this case, the leakage function $\mathcal{L}^{Srch}_\text{Srv}$ is still stateless.
Namely, under this assumption, protocol $\texttt{Bamboo}^*.\textsf{Search}$ has the leakage function
\begin{align*}
\mathcal{L}^{Srch}_\text{Srv}(w)=\mathcal{L}_\text{Srv}^{\prime\prime}(&\textsf{sp}(w),\textsf{TimeDB}(w),\\
&\textsf{DUTime}(w),\textsf{KUHist}(U_\text{now}), num_\text{pad}),
\end{align*}
where $\mathcal{L}^{\prime\prime}_\text{Srv}$ is a stateless function.

For $\mathcal{A}_\text{Thf}$, the leakage functions of protocol $\texttt{Bamboo}^*.\textsf{Search}$ and the key-compromise event \textsf{KeyLeak} can no longer be stateless. If the search result volume of a keyword is an adjustable padding value $x^i$, the thief $\mathcal{A}_\text{Thf}$ can gain a higher probability of guessing - what the client is searching for - than the case where the search result volume always equals $a_\text{max}$. Thus, in $\texttt{Bamboo}^*$, the leakage functions $\mathcal{L}^{Srch}_\text{Thf}$ and $\mathcal{L}^{KeyLeak}_\text{Thf}$ are described as:
\begin{gather*}
\mathcal{L}^{Srch}_\text{Thf}(w)=\overline{\mathcal{L}}_\text{Thf}(num_\text{pad}),\\
\mathcal{L}^{KeyLeak}_\text{Thf}=\overline{\overline{\mathcal{L}}}_\text{Thf}(\textsf{CUHist}(U_\text{now}),\textsf{DUHist}(U_\text{now})),
\end{gather*}
where $\overline{\mathcal{L}}_\text{Thf}$ and $\overline{\overline{\mathcal{L}}}_\text{Thf}$ are two stateful functions.

\section{Implementations and Evaluations}\label{SEC.EXPERIMENT}

We implemented \texttt{Bamboo}, $\texttt{Bamboo}^*$, and three key-updatable DSSE schemes and further compared their performance using a real-world dataset.
Those three baseline schemes are revised from DSSE schemes that have the same level of backward security with \texttt{Bamboo}.
Specifically, the selected schemes are \texttt{MITRA}~\cite{DBLP:conf/ccs/ChamaniPPJ18}, \texttt{Fides}~\cite{DBLP:conf/ccs/BostMO17}, and \texttt{Aura}~\cite{DBLP:conf/ndss/SunSLYSLNG21}.
We denote their corresponding key-updatable versions by $\texttt{MITRA}^\text{KU}$, $\texttt{Fides}^\text{KU}$, and $\texttt{Aura}^\text{KU}$, respectively.
We did not choose $\texttt{SD}_a$ and $\texttt{SD}_d$ for comparison because Sun et al. have proved that \texttt{Aura} outperforms $\texttt{SD}_a$ and $\texttt{SD}_d$~\cite{DBLP:conf/ndss/SunSLYSLNG21}.
The \textsf{KeyUpdate} processes of $\texttt{Fides}^\text{KU}$ and $\texttt{Aura}^\text{KU}$ are interactive.
Namely, the client needs to download, decrypt and re-encrypt, and re-upload the entire encrypted database to update the secret key.
$\texttt{MITRA}^\text{KU}$ was implemented by replacing the PRF function of \texttt{MITRA} with a key-updatable one and encapsulating the file identifier with the mapping function $\pi$.
In this way, $\texttt{MITRA}^\text{KU}$ is equipped with a non-interactive \textsf{KeyUpdate} protocol.
As discussed in Section~\ref{SEC.INTRODUCTION}, $\texttt{Aura}^\text{KU}$, $\texttt{Fides}^\text{KU}$, and $\texttt{MITRA}^\text{KU}$ cannot achieve post-compromise security, since they cannot guarantee the ciphertext security generated during the special time slot.
Note that we did not implement a padding process during the search for the baseline key-updatable DSSE schemes.

\subsection{Experimental Setup}\label{SEC.EXPSETUP}

We used a client and a server connected via a LAN network to perform the experiments.
Table~\ref{TAB.HARDWARE} presents the hardware and operating system configurations of the client and the server, respectively.
They are connected via the Ethernet with about 100 Mbps bandwidth and about a one-millisecond delay.
To yield comprehensive experiments, we additionally created a network environment with about 300 milliseconds delay.
The extra network delay is produced with the ``tc'' command offered by the operating system.

\begin{table}[h]
\centering
\caption{Hardware and OS Configurations.}
\label{TAB.HARDWARE}
\begin{adjustbox}{width=0.8\columnwidth,center}
\begin{tabular}{|c|c|c|}
\hline
 &  Client &  Server\\
\hline
CPU & AMD Ryzen 9 5950X & Intel Xeon Silver 4216 \\
\hline
Memory & 128 GB & 128 GB\\
\hline
Disk Drive  & \multicolumn{2}{c|}{256 GB SAMSUNG PM981 NVME SSD} \\
\hline
OS & \multicolumn{2}{c|}{Ubuntu Server 20.04 x64} \\
\hline
\end{tabular}
\end{adjustbox}
\end{table}

We coded \texttt{Bamboo}, $\texttt{Bamboo}^*$, and the baseline schemes in C++.
All the evaluated schemes use a native TCP socket to establish network communications, SQLite database~\cite{SQLite} as their client states, and PostgreSQL database~\cite{PostgreSQL} to store ciphertexts.
In particular, we used the command ``PRAGMA synchronous=off'' to disable the database synchronization mechanism of SQLite.
Hash functions, PRF functions, and the invertible mapping functions in \texttt{Bamboo}, $\texttt{Bamboo}^*$, and $\texttt{MITRA}^\text{KU}$ are implemented with OpenSSL~\cite{OpenSSL} (which provides SHA-256, SHA-384, and SHA-512 cryptographic hash functions) and Relic Toolkit~\cite{relic-toolkit} (which provides NIST-P256 elliptic curve algorithms).
We used the GMP library~\cite{GMP} to realize the RSA-based permutation that is used in $\texttt{Fides}^\text{KU}$.
$\texttt{Aura}^\text{KU}$ is developed based on the code provided by Sun et al.~\footnote{\url{https://github.com/MonashCybersecurityLab/Aura}}
All implementations can achieve a 128-bit security level.

Our test dataset is extracted from English Wikipedia~\cite{wiki}.
Specifically, we used WikiExtractor~\cite{Wikiextractor2015} to process the Wikipedia~\cite{wiki} and then ran porter stemmer~\cite{DBLP:journals/program/Porter80} to extract keywords from the processed data.
We treated one article as a single document and directly used the identifier number of each article as the file identifier. The length of the file identifier is, at most, 8 bytes.
We also chose some of the extracted data to produce a dataset containing 3,257,613 pairs of keyword and file identifiers.
The dataset contains twenty-five keywords. The number of those twenty-five keywords matching file identifiers ranges from about 10,000 to about 250,000.
Those twenty-five keywords are sufficient to yield a comprehensive evaluation of time costs.
To match with the dataset, we adaptively fine-tuned the parameter of $\texttt{Aura}^\text{KU}$, especially the supported maximum number of deletions $d=150,000$, the false positive rate $p=10^{-5}$, and the hash function number $h=13$ of the underlying Bloom Filter.
We also set the parameter $a_\text{max}=410,000$ of \texttt{Bamboo} and $\texttt{Bamboo}^*$.
In $\texttt{Bamboo}^*$, we set the integer $x=2$, which will be used in function $\textsf{PaddingVal-adj}$.
Note $\texttt{MITRA}^\text{KU}$ and $\texttt{Fides}^\text{KU}$ do not have such parameters for fine-tuning.

In the experiments, we comprehensively tested and compared the performance of \textsf{DataUpdate}, \textsf{KeyUpdate}, and \textsf{Search} of \texttt{Bamboo}, $\texttt{MITRA}^\text{KU}$, $\texttt{Fides}^\text{KU}$, and $\texttt{Aura}^\text{KU}$ using our dataset.
Since the essential difference between \texttt{Bamboo} and $\texttt{Bamboo}^*$ is in the \textsf{Search} protocol, we only evaluated the \textsf{Search} performances of them.
Specifically, when testing \textsf{DataUpdate} performance, we reported those schemes' average client time costs to generate one $add$ \textsf{DataUpdate} query when encrypting the entire database.
We also tested the average client time costs on generating one $del$ \textsf{DataUpdate} query when issuing $d=150,000$ $del$ queries.
In addition, we evaluated the client time costs on reading and then writing the client state when generating a \textsf{DataUpdate} query with $op=del$ or $op=add$.

The \textsf{KeyUpdate} and \textsf{Search} performance are tested in both network environments with delays of about one millisecond and 300 milliseconds, respectively.
While testing the \textsf{KeyUpdate} performance, we encrypted the whole dataset and then updated the encryption key of the generated ciphertexts to evaluate the total time and client time costs for the schemes.
We evaluated the \textsf{Search} performance under two conditions - with and without (document) deletion.
The evaluation metrics are the same as those we used in the experiments for \textsf{KeyUpdate}.
In the examination of the \textsf{Search} performance without deletion, we encrypted the dataset and then performed searches for all keywords.
The experiment for the case with deletion is similar, but with the exception that after encrypting the dataset, we chose a keyword with about 150,000 matching file identifiers and further issued different numbers of \textsf{DataUpdate} queries with $op=del$ of the keyword to the server.
These queries are made up of 0\%-90\% randomly selected file identifiers of the corresponding search results to the keyword.
We then evaluated the \textsf{Search} efficiency on the keyword.

Finally, we compared the client and server storage costs of \texttt{Bamboo}, $\texttt{MITRA}^\text{KU}$, $\texttt{Fides}^\text{KU}$, and $\texttt{Aura}^\text{KU}$ (note $\texttt{Bamboo}$ and $\texttt{Bamboo}^*$ share the same client/server storage complexity).
We extracted additional keywords from English Wikipedia and run \textsf{DataUpdate} with them to test how client-side storage grows with the number of distinct keywords.
We did not report the impacts of file numbers on the client storage as, in the implementations, all counters related to files were fixed to 4-byte integers.
In terms of the server-side storage, we reported the size of PostgreSQL tables under various numbers of $add$ entries (i.e., the entries having $op=add$).

\subsection{\textsf{DataUpdate} Performance}

\begin{table}[h]
\centering
\caption{Comparison on Average Client Time Cost ($\mu$s) of \textsf{DataUpdate}.}
\label{TAB.DUPDT}
\begin{adjustbox}{width=0.85\columnwidth,center}
\begin{tabular}{|c|c|c|c|c|}
\hline
& \texttt{Bamboo} & $\texttt{MITRA}^\text{KU}$ & $\texttt{Aura}^\text{KU}$ & $\texttt{Fides}^\text{KU}$ \\
\hline
 $op=add$ & 1.96$\times 10^3$& 1.08$\times 10^3$& 7.07$\times 10^4$& 6.19$\times 10^3$\\
 \hline
 $op=del$ & 1.96$\times 10^3$& 1.08$\times 10^3$& 7.04$\times 10^4$ & 6.14$\times 10^3$ \\
\hline
\hline
Client State & 59.73 & 54.18 &6.97$\times 10^4$  & 55.21 \\
\hline
\end{tabular}
\end{adjustbox}
\end{table}

We present the evaluation of \textsf{DataUpdate} in Table~\ref{TAB.DUPDT}.
Note that the average \textsf{DataUpdate} time costs of all the compared schemes are constant, and they are not affected by any historical \textsf{DataUpdate}, \textsf{KeyUpdate}, and \textsf{Search} queries.
We can see that \texttt{Bamboo} outperforms $\texttt{Aura}^\text{KU}$ and $\texttt{Fides}^\text{KU}$.
Specifically, \texttt{Bamboo} saves about 97.22\% and 68.33\% client time costs compared to $\texttt{Aura}^\text{KU}$ and $\texttt{Fides}^\text{KU}$.
Table~\ref{TAB.DUPDT} also presents the time costs of accessing the SQLite-based client state when issuing a \textsf{DataUpdate} query.
The client state cost is the same in both cases of $op=add$ and $op=del$.
Over 98\% of $\texttt{Aura}^\text{KU}$'s \textsf{DataUpdate} overhead comes from accessing the client state.
This may indicate the importance of maintaining lightweight access (of client state) in practice.
Fortunately, \texttt{Bamboo} satisfies this requirement.
Say, it only consumes about 59.73 microseconds to access the SQLite database, and the result is very close to the cost of $\texttt{MITRA}^\text{KU}$ and $\texttt{Fides}^\text{KU}$ (in which the performance gap is $< 6$ microseconds).
Compared to $\texttt{MITRA}^\text{KU}$, \texttt{Bamboo} only requires a slight extra cost ($\le$ $8.80\times 10^2$ microseconds) to issue a \textsf{DataUpdate} query.
This is because \texttt{Bamboo} produces one more part in the ciphertext of a single entry (than $\texttt{MITRA}^\text{KU}$) to maintain the chain-link inter-ciphertext structure, which results in an extra exponentiation operation over the elliptic curve element.

\subsection{\textsf{KeyUpdate} Performance}

\begin{table}[h]
\centering
\caption{Comparison on \textsf{KeyUpdate} Performance ($\mu$s).}
\label{TAB.KeyUpdt}
\begin{adjustbox}{width=\columnwidth,center}
\begin{tabular}{|c|c|c|c|c|c|}
\hline
Delay (ms)& & \texttt{Bamboo} & $\texttt{MITRA}^\text{KU}$ & $\texttt{Aura}^\text{KU}$ & $\texttt{Fides}^\text{KU}$ \\
\hline
\multirow{2}{*}{$1$} & Total Time Cost & 6.04$\times 10^9$ & 4.15$\times 10^9$  & 9.06$\times 10^9$& 2.07$\times 10^{10}$ \\
\cline{2-6} & Client Time Cost & 24.49  & 12.93 & 1.34$\times 10^9$ & 2.02$\times 10^{10}$\\
\hline
\multirow{2}{*}{$300$}  & Total Time Cost & 6.04$\times 10^9$ & 4.15$\times 10^9$  & 1.55$\times 10^{10}$ & 2.21$\times 10^{10}$ \\
\cline{2-6}              & Client Time Cost  & 19.20 & 10.90  & 1.36$\times 10^9$  & 2.02$\times 10^{10}$\\
\hline
\end{tabular}
\end{adjustbox}
\end{table}

We show the comparisons on \textsf{KeyUpdate} in Table~\ref{TAB.KeyUpdt}.
\texttt{Bamboo} outperforms $\texttt{Aura}^\text{KU}$ and $\texttt{Fides}^\text{KU}$ in all the metrics.
Specifically, when the network delay is 300 milliseconds, the total time cost of \texttt{Bamboo} to update the key of the encrypted database is only approximately $6.04\times 10^9$ microseconds (about 100 minutes), which is about 3.66 times and 2.57 times faster than those of $\texttt{Fides}^\text{KU}$ and $\texttt{Aura}^\text{KU}$, respectively.
The client time cost of \texttt{Bamboo} is almost negligible compared to the overheads brought by $\texttt{Aura}^\text{KU}$ and $\texttt{Fides}^\text{KU}$.
We notice that the network quality makes less impact on \texttt{Bamboo} than $\texttt{Aura}^\text{KU}$ and $\texttt{Fides}^\text{KU}$.
For example, the absolute difference of \texttt{Bamboo}'s \textsf{KeyUpdate} time costs between the two network environments is about 6.77$\times 10^5$ microseconds (0.67 seconds), while \texttt{Aura} and \texttt{Fides} incur approximately 1.01$\times 10^9$ and 1.41$\times 10^8$ microseconds, respectively.
This is because \texttt{Bamboo}'s client only transfers one small token to the server to execute \textsf{KeyUpdate} while the clients of $\texttt{Aura}^\text{KU}$ and $\texttt{Fides}^\text{KU}$ have to download and re-upload the whole encrypted database.
As the increase of network delay, \texttt{Bamboo} may provide more advantage in \textsf{KeyUpdate} than $\texttt{Aura}^\text{KU}$ and $\texttt{Fides}^\text{KU}$.
\texttt{Bamboo} takes longer than $\texttt{MITRA}^\text{KU}$ in \textsf{KeyUpdate}, because it maintains one more component in a single ciphertext (than $\texttt{MITRA}^\text{KU}$) to achieve both post-compromise security and sub-linear (search) complexity.

\begin{figure}[t]
\centering
\includegraphics[width=0.86\columnwidth]{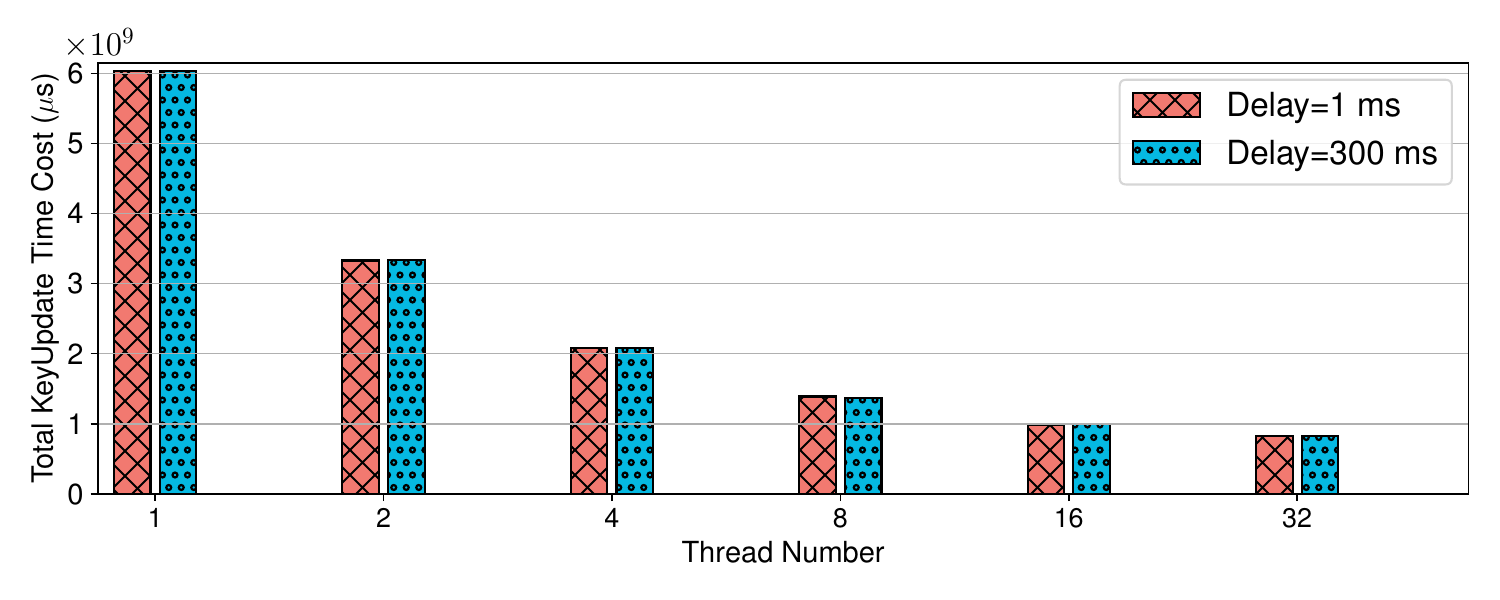}
\caption{Total \textsf{KeyUpdate} Time Cost of \texttt{Bamboo} vs. Number of Threads.}\label{Fig.BambooKeyUpdt}
\end{figure}

The \textsf{KeyUpdate} protocol of \texttt{Bamboo} can be accelerated with multi-thread technique in practice.
Figure~\ref{Fig.BambooKeyUpdt} reports the total time cost of \texttt{Bamboo}'s \textsf{KeyUpdate} when running with different numbers of threads in two network environments.
The results indicate that using multi-thread can help us to reduce the time cost, and they further confirm again that the network quality makes little impact on the \textsf{KeyUpdate} performance.
For example, leveraging sixteen threads, the \textsf{KeyUpdate} only takes nearly $9.80\times 10^8$ microseconds (about 16.33 minutes) to update the key of the whole database in both network environments, which is approximately 6.16 times faster than the single-thread approach.
With this trend, updating the whole database can lead to less cost as the increase of thread number.

\begin{figure}[h]
\begin{minipage}[h]{0.98\linewidth}
\subfigure[Delay=1 ms.]
{
\includegraphics[width=0.46\linewidth]{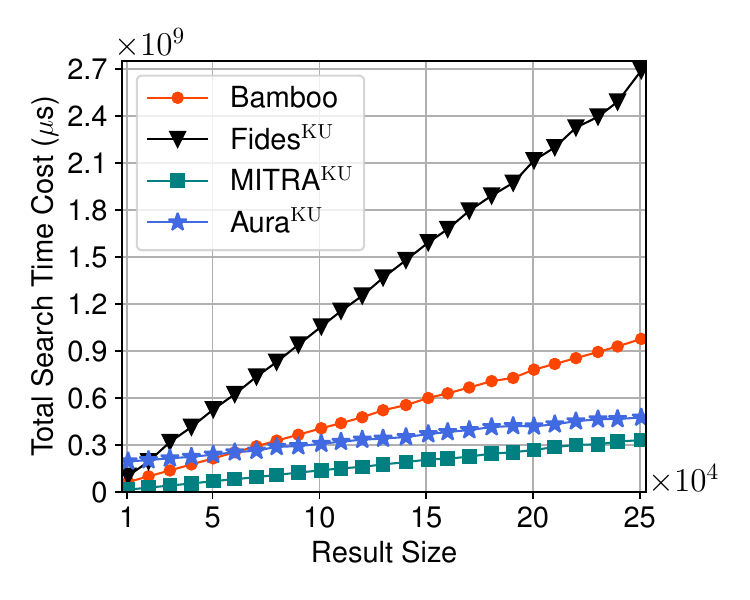}\label{Fig.SrchTimeWithoutDel-1ms}
}
\subfigure[Delay=300 ms.]
{
\includegraphics[width=0.46\linewidth]{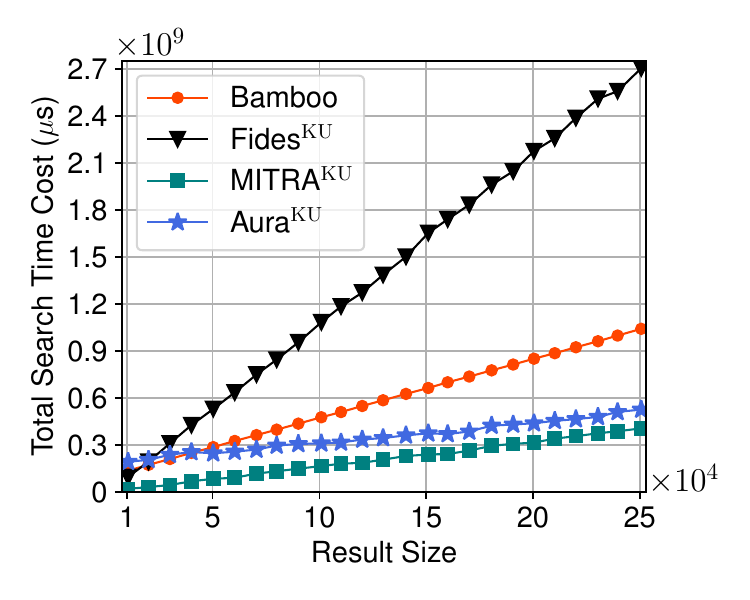}\label{Fig.SrchTimeWithoutDel-300ms}
}
\caption{Total Search Time Cost vs. Result Size without Deletion.}\label{Fig.SrchTimeWithoutDel}
\end{minipage}
\end{figure}

\subsection{\textsf{Search} Performance}

\textbf{\textsf{Search} without Deletion.}
Figure~\ref{Fig.SrchTimeWithoutDel} reports the total search time costs of the evaluated schemes.
The figure clearly shows the linear relationships between the \textsf{Search} time costs and result sizes of all the compared schemes.
We see that when the result size $>20,000$, \texttt{Bamboo} outperforms $\texttt{Fides}^\text{KU}$ in both network environments.
We state that when the result size $\le 20,000$, the performance gap between \texttt{Bamboo} and $\texttt{Fides}^\text{KU}$ is not significant. For example, when the result size is approximately $10,000$ in both network environments, $\texttt{Bamboo}$ costs at most extra 4.20$\times 10^7$ microseconds (about 0.70 minutes) than $\texttt{Fides}^\text{KU}$.
The results show that the network delay makes less impact on \texttt{Bamboo} than $\texttt{MITRA}^\text{KU}$ in the stage of \textsf{Search}.
For example, searching for the keyword corresponding to about $250,000$ matching ciphertexts, the \texttt{Bamboo}'s performance with 300 ms delay is 6.57\% worse than that of the case when Delay=1 ms; and similarly, $\texttt{MITRA}^\text{KU}$ consumes about 22.46\% more cost with 300 ms delay.
We also see that \texttt{Bamboo} is less efficient than $\texttt{Aura}^\text{KU}$.
However, this gap is acceptable.
For example, in both network environments, \texttt{Bamboo} just requires at most an extra $1.82\times 10^3$ microseconds than $\texttt{Aura}^\text{KU}$ to find a matching ciphertext, on average.
\texttt{Bamboo} requires more exponentiation operations over the elliptic curve elements than $\texttt{MITRA}^\text{KU}$ and $\texttt{Aura}^\text{KU}$ during the search.
Those operations, despite their high computational costs, are necessary for the server to search over the chain-like structure correctly.

\begin{figure}[h]
\begin{minipage}[h]{0.98\linewidth}
\centering
\includegraphics[width=0.7\linewidth]{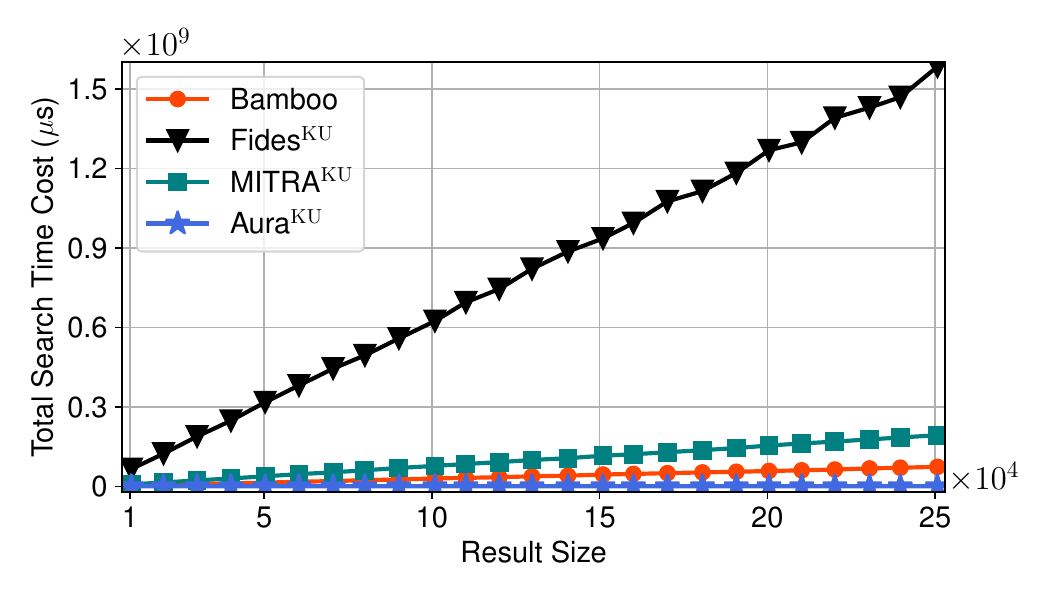}
\caption{Client Search Time Cost vs. Result Size without Deletion.}\label{Fig.ClntTimeWithoutDel}
\end{minipage}
\end{figure}

Figure~\ref{Fig.ClntTimeWithoutDel} presents the experimental results about client time costs on \textsf{Search}.
Since the network delay (no matter how long the delay is) does not affect the cost of the client side, we only present the results when Delay= 1 ms.
In this experiment, one may see that \texttt{Bamboo} performs better than both $\texttt{Fides}^\text{KU}$ and $\texttt{MITRA}^\text{KU}$.
Specifically, when the result size is about 250,000, the cost of \texttt{Bamboo} is only around 7.42$\times 10^7$, saving at least 95.32\% and 61.65\% overheads as compared to $\texttt{Fides}^\text{KU}$ and $\texttt{MITRA}^\text{KU}$, respectively.
If the result size continues increasing, \texttt{Bamboo} will reduce more time compared to $\texttt{Fides}^\text{KU}$ and $\texttt{MITRA}^\text{KU}$.
We notice $\texttt{Aura}^\text{KU}$ has a small advantage over \texttt{Bamboo}, as the \texttt{Bamboo}'s client has to perform decryption to obtain the final results.

\begin{figure}[h]
\begin{minipage}[ht]{0.98\linewidth}
\subfigure[Delay=1 ms.]
{
\includegraphics[width=0.46\linewidth]{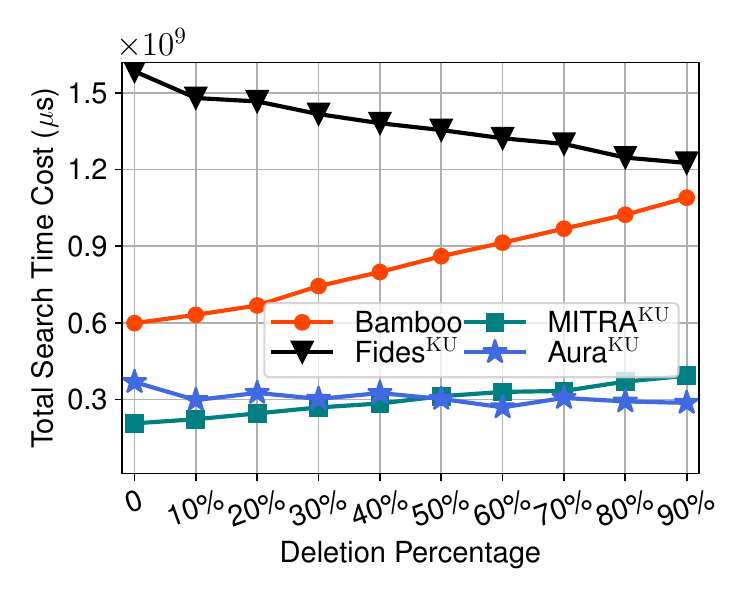}\label{Fig.SrchTimeWithDel-1ms}
}
\subfigure[Delay=300 ms.]
{
\includegraphics[width=0.46\linewidth]{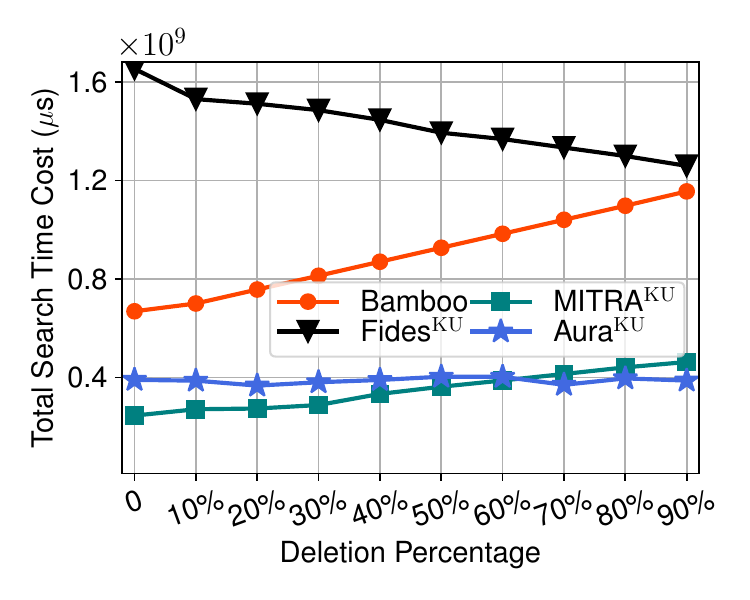}\label{Fig.SrchTimeWithDel-300ms}
}
\caption{Total Search Time Cost vs. Percentage of Deletion.}\label{Fig.SrchTimeWithDel}
\end{minipage}
\end{figure}

\textbf{\textsf{Search} with Deletion.}
Figure~\ref{Fig.SrchTimeWithDel} shows the total search time cost as the deletion percentage varies in the network environments.
From the results, we can conclude that network delay does not significantly affect the \textsf{Search} performance of \texttt{Bamboo} when there are historical deletion queries.
Specifically, \texttt{Bamboo} has an overhead of 8.54$\times 10^7$ microseconds (1.42 minutes) to complete the \textsf{Search} in Figure~\ref{Fig.SrchTimeWithDel-300ms} as compared to Figure~\ref{Fig.SrchTimeWithDel-1ms} w.r.t. the same keyword.
With the increase of the deletion percentage, the cost of $\texttt{Fides}^\text{KU}$ decreases.
The reason is that $\texttt{Fides}^\text{KU}$'s client needs to re-encrypt and re-upload fewer ciphertexts during \textsf{Search} (as the deletion number increases).
Even so, in general, \texttt{Bamboo} still outperforms $\texttt{Fides}^\text{KU}$.
For example, when the percentage is set to 40\% in Figure~\ref{Fig.SrchTimeWithDel-300ms}, \texttt{Bamboo} consumes around 8.70$\times 10^8$ microseconds in total, which reduces 39.84\% cost compared with $\texttt{Fides}^\text{KU}$.
Even in the worst case, say, the percentage=90\% in Figure~\ref{Fig.SrchTimeWithDel-300ms}, \texttt{Bamboo} costs at most another $5.09\times 10^4$ microseconds, compared with others, to find a matching ciphertext.
This performance gap exists due to the same reason explained in the ``search without deletion".

\begin{figure}[h]
\begin{minipage}[ht]{0.98\linewidth}
\centering
\includegraphics[width=0.7\linewidth]{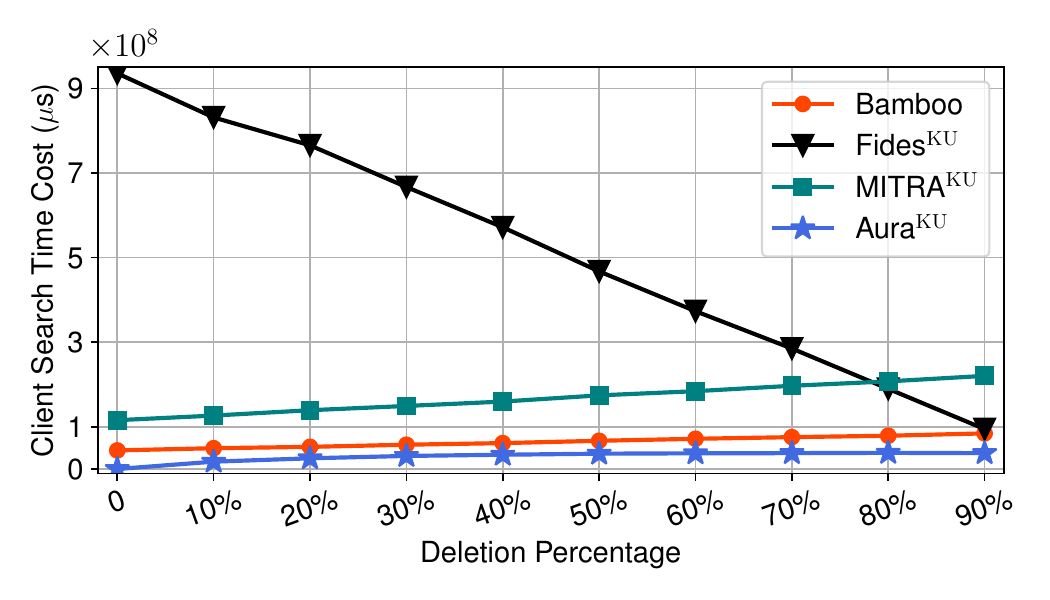}
\caption{Client Time Cost vs. Percentage of Deletion.}\label{Fig.ClntTimeWithDel}
\end{minipage}
\end{figure}

Figure~\ref{Fig.ClntTimeWithDel} reports the cost on the client side with the change in deletion percentage.
Since the network delay does not affect the client cost, we only present the results when Delay=1 ms.
\texttt{Bamboo} keeps its advantages over $\texttt{Fides}^\text{KU}$ and $\texttt{MITRA}^\text{KU}$.
This indicates that \texttt{Bamboo}'s \textsf{Search} is cost-effective and computation friendly to the client in general.
When the percentage is 40\% in Figure~\ref{Fig.ClntTimeWithDel}, our client only takes about 6.19$\times 10^7$ microseconds (roughly 1.03 minutes) in \textsf{Search}, approximately 9.24 times and 2.59 times more efficient than $\texttt{Fides}^\text{KU}$ and $\texttt{MITRA}^\text{KU}$, respectively.
One may also see that the gap between \texttt{Bamboo} and $\texttt{Aura}^\text{KU}$ is actually quite close.
For example, when the percentage=90\% \texttt{Bamboo} take $3.09\times 10^3$ microseconds more to locate a matching ciphertext than $\texttt{Aura}^\text{KU}$.
This is because \texttt{Bamboo} requires the client to perform decryption and filter out the deleted file identifiers to obtain the search results, while $\texttt{Aura}^\text{KU}$ does not.

\begin{figure}[h]
\centering
\begin{minipage}[ht]{0.98\linewidth}
\subfigure[Total Search Time Cost vs. Result Size without Deletion.]
{
\includegraphics[width=0.46\linewidth]{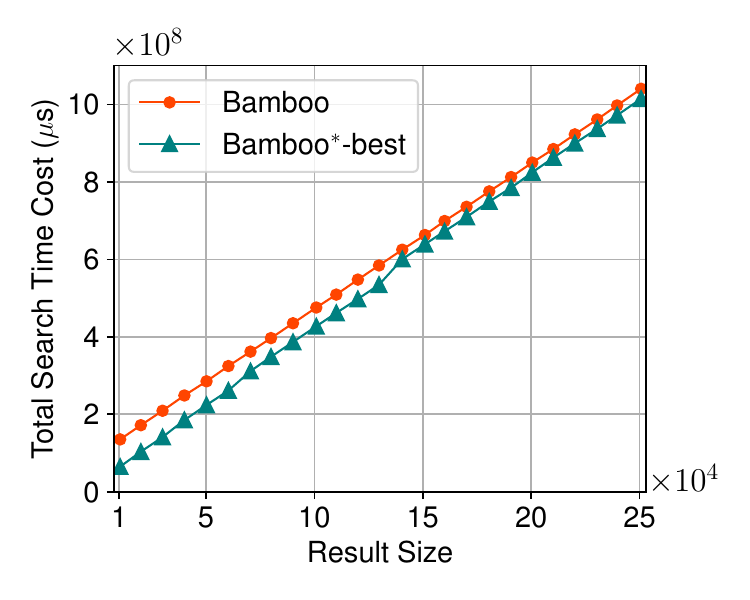}\label{Fig.Bamboo-star-without-del}
}
\subfigure[Total Search Time Cost vs. Percentage of Deletion.]
{
\includegraphics[width=0.46\linewidth]{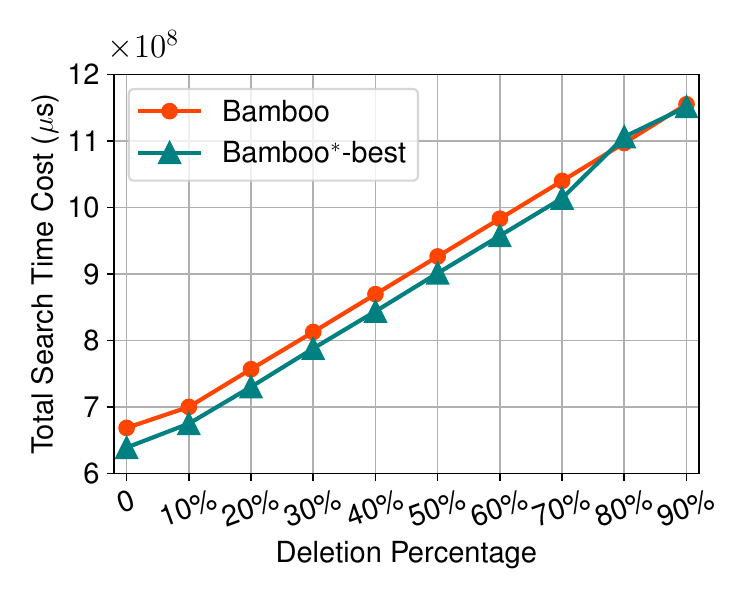}\label{Fig.Bamboo-star-with-del}
}
\caption{\textsf{Search} Comparison \texttt{Bamboo} vs. $\texttt{Bamboo}^*$.}\label{Fig.Bamboo-star}
\end{minipage}
\end{figure}

\textbf{\textsf{Search} Comparison between \texttt{Bamboo} and $\texttt{Bamboo}^*$.}
In this part, we compare the \textsf{Search} Performance with and without historical deletions between \texttt{Bamboo} and $\texttt{Bamboo}^*$.
For the case when Delay=1 ms, the performance gap between the two schemes is not significant.
Therefore, we only present the experimental results when Delay=300 ms.
For $\texttt{Bamboo}^*$, we tested and recorded the best case of its \textsf{Search}.
This case happens when the function \textsf{PaddingVal-adj} returns the adjustable padding value (instead of the maximum padding one) if the padding value could be used.
We note that the worst case is the other way round, i.e., returning the maximum padding value, which is equal to the $\texttt{Bamboo}$'s \textsf{Search}.
We name the best case $\texttt{Bamboo}^*$-best.
From Figure~\ref{Fig.Bamboo-star}, we see that the smaller the number of matching ciphertexts we have, the more time, in the $\texttt{Bamboo}^*$-best, we save as compared to \texttt{Bamboo}.
We also notice that in both sub-figures, the gap of both lines is shrinking as the increase of result size and deletion percentage.
For instance, in Figure~\ref{Fig.Bamboo-star-without-del}, when the result size is about 10,000, $\texttt{Bamboo}^*$-best saves about 70.32 seconds over \texttt{Bamboo} to complete the \textsf{Search}; whilst the size reaches about 250,000, the advantage decreases to 25.67 seconds.
Similarly, assuming the deletion percentage=30\% (namely, there are in total 105,697 matching ciphertexts), the cost we save is about 24.80 seconds in Figure~\ref{Fig.Bamboo-star-with-del}. Then, if the percentage is greater than 80\%, the gap disappears. This is because when the percentage is greater than 80\%, the adjustable padding value is greater than the maximum padding value, and thus the maximum padding value is used to complete the padding process.

\subsection{Storage Efficiency}\label{SEC.EXPSTOR}

\begin{figure}[h]
\centering
\begin{minipage}[t]{0.46\linewidth}
\centering
\includegraphics[width=\linewidth]{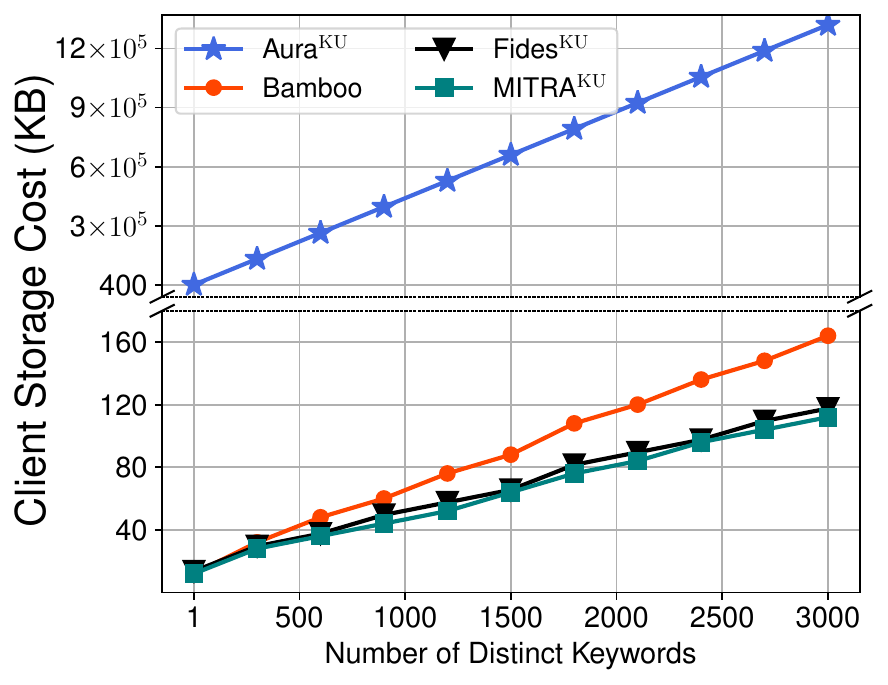}
\caption{Client Storage Cost vs. Number of Distinct Keywords}\label{Fig.Clnt-Stor}
\end{minipage}
\begin{minipage}[t]{0.495\linewidth}
\centering
\includegraphics[width=\linewidth]{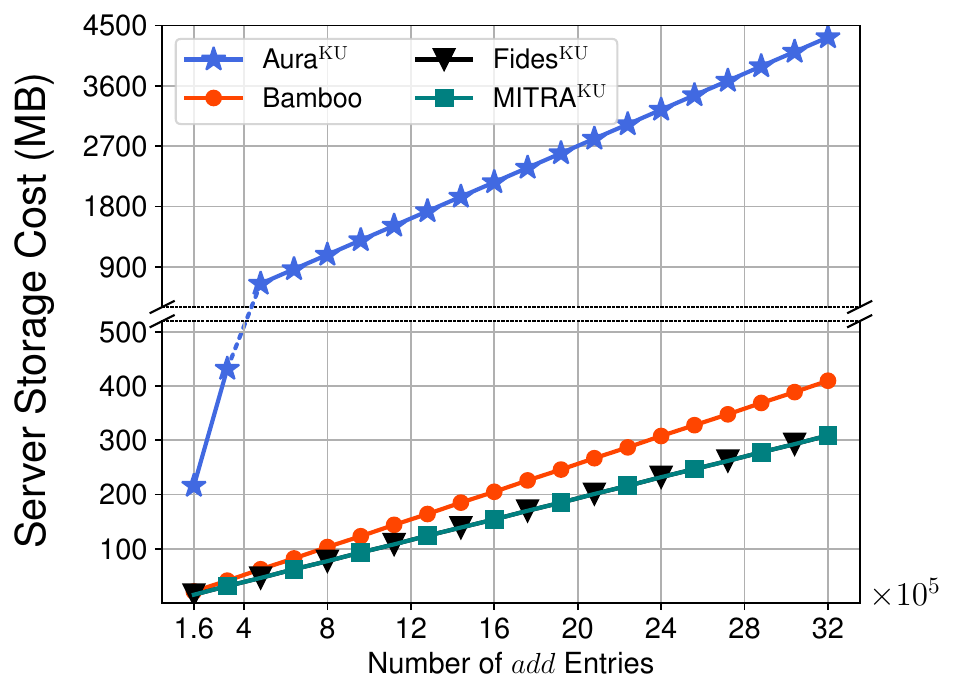}
\caption{Server Storage Cost vs. Number of $add$ Entries}\label{Fig.Srv-Stor}
\end{minipage}
\end{figure}

Figure \ref{Fig.Clnt-Stor} presents a comparison of the client-side storage costs.
\texttt{Bamboo} clearly outperforms $\texttt{Aura}^\text{KU}$.
When the client executes \textsf{DataUpdate} with 2,100 distinct keywords, \texttt{Bamboo} only requires roughly 120.06 KB data, saving 99.98\% storage as compared to $\texttt{Aura}^\text{KU}$.
This is because \texttt{Bamboo} only records a $\lambda-$bit random number in the private state and a 4-byte counter for each keyword, but $\texttt{Aura}^\text{KU}$ applies a Bloom Filter (with tens of thousands of bits).
The \texttt{Bamboo}'s client stores more data than $\texttt{MITRA}^\text{KU}$ and $\texttt{Fides}^\text{KU}$. 
That extra overhead is acceptable in practice.
For example, given 3,000 distinct keywords, \texttt{Bamboo} takes about 164.06 KB, in which there are only $\le52.03$ KB extra costs as compared to $\texttt{MITRA}^\text{KU}$ and $\texttt{Fides}^\text{KU}$.

Figure~\ref{Fig.Srv-Stor} shows the server storage costs under various numbers of $add$ entries.
We see that \texttt{Bamboo} surpasses $\texttt{Aura}^\text{KU}$. 
For instance, given 2,400,000 $add$ entries, the \texttt{Bamboo}'s server consumes 308 MB storage, which is about 9.5\% of the cost of $\texttt{Aura}^\text{KU}$ (3,242 MB).
Compared to $\texttt{MITRA}^\text{KU}$ and $\texttt{Fides}^\text{KU}$, 
the \texttt{Bamboo}'s ciphertext of a single ($add$) entry yields an extra part. 
This additional cost is minor.
Assuming the encrypted database contains 3,200,000 entries, the cost is only 101 MB on the server side. 

As a conclusion, considering the total cost in \textsf{KeyUpdate} and \textsf{DataUpdate}, \texttt{Bamboo} outperforms $\texttt{Fides}^\text{KU}$ and $\texttt{Aura}^\text{KU}$ while achieving similar performance to $\texttt{MITRA}^\text{KU}$.
It is interesting to see that \textsf{KeyUpdate} of \texttt{Bamboo} can be accelerated via the multi-thread technique.
In terms of \textsf{Search} efficiency, \texttt{Bamboo} maintains the same level of practicability as those baseline schemes.
In addition, \texttt{Bamboo} gains noticeable advantages on the client side among those schemes in terms of the costs in \textsf{DataUpdate}, \textsf{KeyUpdate}, and \textsf{Search}.
Moreover, \texttt{Bamboo} achieves practical storage performance, which is very close to that of $\texttt{MITRA}^\text{KU}$ and $\texttt{Fides}^\text{KU}$, on both the client and server sides.
Given that \texttt{Bamboo} captures the post-compromise security, we can conclude that it is the first practical DSSE scheme with high performance and strong security in the literature.
As an improved variant, $\texttt{Bamboo}^*$ can save \textsf{Search} time cost. 
Moreover, it is applicable to those large-scale databases where there exist many keywords which correspond to a small amount of matching file identifiers, such as the English Wikipedia.
We provide more discussions on the above experiments in Section~\ref{Diss.Expr}.

\section{Discussions and Future Works}

\subsection{Discussions on Trivial Extension to SEKU}\label{Dis.TrivialExtToSEKU}

Some of the existing DSSE schemes (e.g., \texttt{MITRA}, \texttt{ORION}, \texttt{HORUS}~\cite{DBLP:conf/ccs/ChamaniPPJ18}) can be extended to support \textsf{KeyUpdate} by replacing their PRF functions or symmetric encryption schemes with key-updatable PRF~\cite{DBLP:conf/crypto/BonehLMR13,9724186} or updatable encryption~\cite{DBLP:conf/eurocrypt/LehmannT18,DBLP:conf/eurocrypt/KloossLR19,DBLP:conf/crypto/BoydDGJ20,DBLP:conf/asiacrypt/Jiang20}.
However, such an extension may not achieve post-compromise security.
As we have pointed out in Section~\ref{SEC.INTRODUCTION}, they cannot guarantee the security of ciphertexts generated in the special time slot (i.e., after the key compromise before the \textsf{KeyUpdate}).
This is because they cannot provide enough \emph{unpredictable private randomness} to generate a ciphertext.
Private randomness guarantees that the randomnesses (e.g., the honest, randomly generated secret key) should be only known to the client.
In each of the existing schemes, the \textsf{DataUpdate} protocol generates one or more parts of the ciphertext with only static or derivable private randomness, e.g., a secret key of a CPA-secure encryption scheme or a secret key derived from a counter.
Once the secret key and the private state are exposed, the thief can easily extract information from the corresponding parts of the ciphertexts within the special time slot.
From the above discussions, we conclude that simply extending existing schemes may not provide a post-compromise-secure solution.

\subsection{Discussions on Type 2 Threat Model}\label{Diss.DiscussType2Model}

Following the philosophy of DSSE, we say that a server should be honest-but-curious (i.e., under Type 1 model).
But it becomes extremely powerful in Type 2 model.
This yields further impossibilities in the design.
A natural concern is how we could detect and resist data injection attacks launched by the server.
Given the compromised secret key, the server can easily inject and tamper with the data in the database.
All changes made by the server are now essentially ``valid" due to the knowledge of the key.
It could be feasible for the client to locally maintain extra verification information of the encrypted database so as to detect any illegal operations on the database, like~\cite{DBLP:journals/iacr/BostFP16}.
This approach may significantly increase client-side storage, computation, and communication costs.
Further, it is required that the verification information should be stored separately from the secret key on the client side.
And so far there is no evidence that this verifiable approach is practically secure.
For example, the server is able to obtain the state information (including the verification knowledge, e.g., how the verification is done) of the encrypted database with the compromised key.
With that knowledge, the server may adaptively perform malicious operations which can bypass the verification.

Another challenge is to guarantee the security of the keyword search after a key compromise.
Once the key compromise happens, the server learns the exact search results and frequencies of all the keywords stored in the encrypted database.
The leaked information can be exploited to infer the underlying keyword of the subsequent client's keyword queries~\cite{DBLP:conf/uss/OyaK21,DBLP:conf/ccs/CashGPR15,DBLP:conf/ccs/NingHPYL0D21}, even after the \textsf{KeyUpdate}.
Recall that there is a special time slot, the period after the key compromise and before the key update.
In this slot, we still need to protect the ciphertexts and keyword search queries.
The techniques used to design volume-hiding structured encryption~\cite{DBLP:conf/eurocrypt/KamaraM19} and query-equality-suppressing structured encryption~\cite{DBLP:conf/crypto/KamaraMO18, DBLP:conf/eurocrypt/GeorgeKM21} may be the potential solutions.
But it is unknown if it is possible to apply them to the oblivious \textsf{KeyUpdate}.
How to design a secure and practical scheme in Type 2 threat model is an interesting problem.

\subsection{Discussions on \texttt{Bamboo} Construction}
\label{Diss.BambooCons}

\textbf{File Identifier Length.} The construction of \texttt{Bamboo} relies on the DDH assumption, and the file identifier is encoded into a cyclic group element.
This may restrict the file identifier length.
Fortunately, this problem is quite easy to solve.
Specifically, to support a long file identifier, one can split the identifier into small pieces so that each piece can fit the length limitation of a group element.
Those pieces are encoded into elements and then can be respectively encrypted with the same random number and secret key and different hash functions.
After running \texttt{Bamboo}.\textsf{DataUpdate}, those encrypted elements are viewed as a whole file identifier ciphertext and uploaded to the server.
Later, the client can decrypt all the elements and merge the pieces to recover the identifier.
It is easy to see that the above approach does not affect security.

\textbf{File Deletion.} When handling a deletion request on a pair of keyword and file identifier, the \textsf{DataUpdate} may encrypt the operation type $op=del$ with the pair to generate a ciphertext as a special deletion query.
This approach logically marks the pair as ``deleted" but does not remove it from $\mathbf{EDB}$.
This may not be a ``completed" deletion for the pair.
A similar method is used in some existing DSSE, e.g., MITRA.
The reason behind the design is that \textsf{DataUpdate} should not leak anything in the setting of post-compromise security.
We also may not employ the \textsf{Search} to remove the deletion (unlike \texttt{Aura}) since any two \textsf{Search} queries must be indistinguishable in the view of the thief.
A possible enhancement could be to enable \textsf{KeyUpdate} to locate and remove the ciphertexts and further update the key of the remaining ciphertexts.
However, it may be challenging to reduce the information leakage to capture the post-compromise security in this context fully.
We leave this challenge for future research.

\textbf{\textsf{KeyUpdate} Intervals.} Theoretically, the more we execute \textsf{KeyUpdate}, the better we achieve key-compromise security.
In practice, we may use three strategies to balance \textsf{KeyUpdate} and security:
(1) follow the suggestions given by standards, e.g., NIST Special Publication 800-57~\cite{NIST800-57} - the symmetric data-encryption key should be updated within 1-2 years after being created;
(2) when the client detects/suspects the secret key is (partially) leaked;
(3) when the encrypted database stays idle for a certain period.
In fact, the \textsf{KeyUpdate} overhead is practical.
For example, updating the key over the database with about 3,200,000 entries (using two threads) costs roughly 50 minutes.
This overhead is mostly on the server side, while the client needs less than $30$ microseconds.

\textbf{Extension to Type 2 Threat Model.} \texttt{Bamboo} is provably secure under Type 1 threat model.
Unfortunately, it is so far impossible to extend it in Type 2 threat model efficiently.
As explained in Section~\ref{Diss.DiscussType2Model}, the security under Type 2 threat model relies on the robust verification mechanism and oblivious \textsf{KeyUpdate}.
The difficulty is in the design of the latter.
%
%
The current design of \textsf{KeyUpdate} cannot provide obliviousness since the server uses the \textsf{KeyUpdate} token to update the encrypted database ciphertext by ciphertext.
A straightforward solution is to require the client to download, decrypt, re-encrypt, and re-upload the whole database.
This may be extremely costly and does not scale well in practice. A practical oblivious \textsf{KeyUpdate} without strong security assumptions (e.g., relying on a trusted third party) and expensive costs remains an open question.

\textbf{Extension to Multi-Keyword Search.}
SEKU is formalized based on the classic DSSE definition under the single-keyword search context.
Fortunately, we can extend \texttt{Bamboo} to support multi-keyword conjunctive search by a \emph{cross-tag}  technique~\cite{DBLP:conf/crypto/CashJJKRS13,DBLP:conf/ccs/LaiPSLMSSLZ18,DBLP:conf/ndss/PatranabisM21}.
We use two types of encrypted databases on the server side.
One is the traditional encrypted database as in DSSE (named \textbf{TSet}), and the other is to verify the conjunctive relationships between two given keywords (named \textbf{XSet}).
In a $\textsf{DataUpdate}$ with an entry $(op,((w_1,w_2,...,w_k),id))$, we add/delete the records $(w_1,id),(w_2,id),...,(w_k,id)$ to/from \textbf{TSet}, and further construct special tags of the entry to store in \textbf{XSet}.
Whilst handling a conjunctive query $w_1\wedge w_2\wedge...\wedge w_n$, we query \textbf{TSet} to get the file identifiers to $w_1$ and then use \textbf{XSet} to verify if each of the returned identifiers contains $w_2,w_3,..,w_n$.
One may follow the above approach to instantiate \textbf{TSet} with \texttt{Bamboo} and design a post-compromise-secure \textbf{XSet}.

\textbf{Against Inference Attacks.} In the context of key compromise, provided that the encrypted database can be exposed to the thief, one may ask if \texttt{Bamboo} is vulnerable to the inference attacks~\cite{DBLP:conf/ndss/IslamKK12,DBLP:conf/ccs/CashGPR15,DBLP:conf/ndss/BlackstoneKM20,DBLP:conf/uss/OyaK21}.
This type of attack should leverage a sufficient amount of the leakage from the \textsf{Search}.
From the thief's perspective, we say that it cannot see the \textsf{Search} leakage as the client and server communicate via a secure channel by DH exchange; meanwhile, it does not collude with the server.
But if the server actively launches the attack, \texttt{Bamboo} may not perform well, as prior forward and Type-II backward secure DSSE schemes (e.g., \texttt{Fides}~\cite{DBLP:conf/ccs/BostMO17}, \texttt{MITRA}~\cite{DBLP:conf/ccs/ChamaniPPJ18}, \texttt{Aura}~\cite{DBLP:conf/ndss/SunSLYSLNG21}).
We note that they leak the same amount of information to the server during \textsf{Search}.
Existing practical DSSE schemes with backward security also cannot counter the inference attack, as they (by definition) leak the search results and (partial) access pattern to the server.
To mitigate the attack, we may straightforwardly apply a current countermeasure to \texttt{Bamboo}, e.g., volume hiding solution~\cite{DBLP:conf/eurocrypt/KamaraM19} and leakage suppression strategy~\cite{DBLP:conf/eurocrypt/GeorgeKM21}, which may produce extra overhead during both \textsf{DataUpdate} and \textsf{Search} on the client side.

\textbf{Multiple Clients}. Some research works~\cite{DBLP:conf/esorics/SunLSSY16,DBLP:conf/pkc/HamlinSWW18,DBLP:conf/uss/0001C22} enable multiple clients to collaboratively write/read an encrypted database with fine-grained access control.
It is non-trivial for Bamboo to support this.
The main challenge is in modeling security.
Given multiple data owners/users, the role of the thief and all the cases of ``key stealing" should be carefully defined.
For example, the thief could be among the clients with only write permissions, or it compromises those with both write and read rights; and it may further collude other clients to collect sensitive information from the database.
We leave the multi-client case as an open problem.

\subsection{Discussions on Experiments}\label{Diss.Expr}

\textbf{Other Databases.} In the experiment, we used the Postgresql database to store the generated ciphertexts for the compared schemes.
One may choose to use other databases, e.g., MySQL or MongoDB.
We state that the experimental results while using other databases could be slightly different, because they may lead to different overheads when accessing the ciphertexts.
We leave this to the interested readers.

\textbf{Keywords.} One may argue that the 25 keywords used in the test dataset may not produce comprehensive experiments.
In fact, the total size of the test dataset is 3,257,613, and the keywords we used can sufficiently show the performance differences among the compared schemes.
According to the performance variation tendency, we state that using more keywords will not change the conclusion of our current experiments.

\textbf{Network.}
We argue that the real-world network environment is usually influenced by many uncontrollable and unforeseen factors, such as burst network traffics or relay router failure.
Those factors may make the network hard to set up a stable reproducible experimental foundation providing a fair comparison among the schemes.
The experimental network environment was simulated over a stable LAN network, the network delay was artificially produced via the ``tc'' command, and all tested elements can be under our control.
Thus, the simulated network is more beneficial for us to create a relatively fair test environment.

\section{DSSE Revisited and Related Works}\label{SEC.RELATED-WORK}

Kamara et al.~\cite{DBLP:conf/ccs/KamaraPR12} formally defined the syntax and adaptive security for DSSE.
The security concentrates on the information leakage~\cite{DBLP:conf/ccs/CurtmolaGKO06} revealed to the server.
Since then, many DSSE schemes have been proposed to achieve high search efficiency~\cite{DBLP:conf/fc/KamaraP13,DBLP:conf/ccs/HahnK14}, supporting scalable database~\cite{DBLP:conf/ndss/CashJJJKRS14}, physical deletion~\cite{DBLP:conf/acisp/XuL0SWJ17}, and retaining small leakage~\cite{DBLP:conf/ndss/StefanovPS14}.

Zhang et al.~\cite{DBLP:conf/uss/ZhangKP16} proposed the well-known file-injection attack against DSSE.
This attack enables the adversary to actively inject crafted files into the encrypted database to infer the underlying keywords of search queries.
As an effective countermeasure to this attack, forward security has been considered as an essential property, which requires that a newly updated ciphertext leaks nothing about its underlying keyword.
The first forward secure DSSE scheme was proposed by Chang et al.~\cite{DBLP:conf/acns/ChangM05}.
Later, Stefanov et al.~\cite{DBLP:conf/ndss/StefanovPS14} formalized the forward security using the leakage functions.
After that, many forward secure DSSE schemes have been constructed to deliver sub-linear search complexity~\cite{DBLP:conf/ccs/Bost16}, small leakage~\cite{DBLP:conf/crypto/GargMP16}, high practical performance~\cite{DBLP:conf/ccs/KimKLPK17}, and high I/O efficiency~\cite{DBLP:journals/tdsc/SongDYXZ20}.

Another important feature of DSSE, called backward security, was defined by Stefanov et al.~\cite{DBLP:conf/ndss/StefanovPS14}.
It restricts the information leakage about deleted ciphertexts during search queries.
Bost et al.~\cite{DBLP:conf/ccs/BostMO17} formalized three types of backward security with leakage functions.
Since then, many research works have proposed forward and backward secure constructions to achieve small search leakage~\cite{DBLP:conf/ccs/ChamaniPPJ18},\cite{DBLP:conf/esorics/ZuoSLSP19,DBLP:journals/tdsc/LiHWLLDL21}, robustness under fault operations~\cite{9724186}, constant client storage~\cite{DBLP:conf/ndss/DemertzisCPP20,DBLP:journals/tifs/HeCZDX21}, and practical search performance~\cite{DBLP:conf/ccs/SunYLSSVN18,DBLP:conf/ndss/SunSLYSLNG21,DBLP:conf/esorics/ChenXWZSJ21,DBLP:conf/uss/ChamaniPMD22}.

There are other works on searchable symmetric encryption,  e.g., using trusted hardware to reduce the leakage from the  server~\cite{DBLP:conf/eurosec/AmjadKM19,DBLP:conf/sp/MishraPCCP18}, improving the I/O performance of the encrypted database~\cite{DBLP:conf/ndss/MiersM17,DBLP:conf/crypto/DemertzisPP18,DBLP:conf/crypto/BossuatBFMR21}, and enabling conjunctive search~\cite{DBLP:journals/vldb/WuL19,DBLP:journals/iacr/ZuoSLSPW20,DBLP:conf/ndss/PatranabisM21,DBLP:conf/ccs/LaiPSLMSSLZ18}.

All the aforementioned works have an implicit but unrealistic security assumption that the client's secret key will not be compromised.
Once this assumption does not hold, all prior schemes become insecure.
No prior works systematically investigates the key compromise problem and the countermeasures.
This paper contributes to this line of research by developing post-compromise security for DSSE.

\section{Conclusions}

We investigated and initialized the research topic of DSSE with \textsf{KeyUpdate}.
We defined the notion SEKU and formulated the post-compromise security against Type 1 model.
We further constructed the first scheme of its type, the post-compromise-secure instantiation \texttt{Bamboo}, and meanwhile proved its security.
We state that the post-compromise feature may be a practical consideration for real-world applications.
For example, the client may temporarily use a third-party device (e.g., a public computer) to query the encrypted database.
This may risk the exposure of the secret key.
\texttt{Bamboo} may provide an accountable solution that existing DSSE schemes cannot.
As for efficiency, \texttt{Bamboo} can achieve sub-linear search complexity and constant client time cost.
Finally, we evaluated \texttt{Bamboo} with a real-world dataset.
The experimental results show that \texttt{Bamboo} achieves a comparable search performance to the well-studied forward-and-backward secure DSSE schemes and offers high performance in \textsf{KeyUpdate} and client complexity.
To further improve bandwidth, we introduced a flexible padding technique and then leveraged it to construct $\texttt{Bamboo}^*$, which significantly outperforms \texttt{Bamboo}, especially in a large-scale database where there are many keywords with a small size of search results.

\noindent {\bf Acknowledgements}. We would like to thank the shepherd Prof. Gang Qu and the anonymous reviewers for their valuable comments.
This work was partly supported by the National Key Research and Development Program of China under Grant No. 2021YFB3101304, the Wuhan Applied Foundational Frontier Project under Grant No. 2020010601012188, the National Natural Science Foundation of China under Grant No. 62272186 and No. 61872412, and the Guangdong Provincial Key Research and Development Plan Project under Grant No. 2019B010139001.
This work was also supported by EU Horizon research and innovation programme under grant agreement No. 952697 (ASSURED), No. 101021727 (IRIS) and No. 101070052 (TANGO).



\bibliographystyle{IEEEtranS}
\bibliography{reference-shrink}

\appendices

\section{DDH Assumption}\label{APP.DDH}

\begin{Definition}[Decisional Diffie-Hellman (DDH) Assumption]
Let $\mathbb{G}$ be a multiplicative cyclic group of prime order $q$, and $g$ is a generator of $\mathbb{G}$ where $q$ is of $\lambda$ bit-length. We say that the DDH assumption holds in $\mathbb{G}$ if for any PPT adversary $\mathcal{A}$ the probability that $\mathcal{A}$ distinguishes between tuples $(g,g^a,g^b,g^{ab})$ and $(g,g^a,g^b,g^c)$ is negligible in $\lambda$ where $(a,b,c)\overset{\$}{\leftarrow}\mathbb{Z}^*_q\times \mathbb{Z}^*_q\times \mathbb{Z}^*_q$.
\end{Definition}

\section{Proof of Theorem~\ref{THEO.SEKU-SEC}}\label{SEC.SEKU-PROOF}

To prove the post-compromise security against the Type 1 model, we show that \texttt{Bamboo} is $\mathcal{L}_\text{Srv}$-adaptively and $\mathcal{L}_\text{Stl}$-adaptively secure with the corresponding leakage functions defined in Theorem~\ref{THEO.SEKU-SEC}, respectively. 

\begin{algorithm*}[h]
\algsetup{linenosize=\footnotesize} \footnotesize
\caption{Construction of Simulator $\mathcal{S}$.}\label{SIM.SEMI}
\begin{multicols}{2}
\leftline{\underline{$\mathcal{S}.\textsf{Setup}(\mathcal{L}^{Stp}_\text{Srv}(\lambda,a_\text{max})=(\lambda,a_\text{max}))$}}
\begin{algorithmic}[1]
\STATE Initialize an invertible mapping function $(\lambda,\mathbb{G},q,\pi,$ $\pi^{-1})\leftarrow\mathbf{PGen}(\lambda,\ \lambda)$
\STATE Initialize the Diffie-Hellman key exchange protocol parameters including the elliptic curve group $\mathbb{G}^\prime$ of prime order $q^\prime$ and a generator $g^\prime\in\mathbb{G}^\prime$
\STATE Initialize two empty maps $\mathbf{SHist}$ and $\mathbf{EDB}$
\STATE Initialize an empty list $\mathbf{CList}$
\STATE Initialize timestamp $u\leftarrow 0$
\STATE Initialize a search key $K_1\overset{\$}{\leftarrow} \mathbb{Z}^*_q$
\STATE Send $\mathbf{EDB}$ to the server
\end{algorithmic}

\leftline{\underline{$\mathcal{S}.\textsf{DataUpdate}(\mathcal{L}^{DaUpdt}_\text{Srv}(op,(w,id))=\text{NULL})$}}
\begin{algorithmic}[1]
\STATE Accumulate timestamp $u\leftarrow u+1$
\STATE Randomly select three elements $(L_0,D_0,C_0)\overset{\$}{\leftarrow}\mathbb{G}\times\mathbb{G}\times\mathbb{G}$
\STATE Insert record $(u,L_0,D_0,C_0)$ into $\mathbf{CList}$
\STATE Send the simulated ciphertext $(L^{K_1}_0,D^{K_1}_0,C^{K_1}_0)$ to the server
\end{algorithmic}

\leftline{\underline{$\mathcal{S}.\textsf{Search}(\mathcal{L}^{Srch}_\text{Srv}(w)=(\textsf{sp}(w),\textsf{TimeDB}(w),\textsf{DUTime}(w),\textsf{KUHist}(U_\text{now})))$}}
\begin{algorithmic}[1]
\STATE Establish a temporary secure channel with the server using the Diffie-Hellman key exchange protocol
\STATE Accumulate timestamp $u\leftarrow u+1$
\STATE Abort if $\textsf{DUTime}(w)$ is empty
\STATE Retrieve the minimum timestamp $u^\text{srch}_\text{min}$ from $\textsf{sp}(w)$
\STATE Retrieve record $(u_{s},H_L,H_D,H_C)\leftarrow\mathbf{SHist}[u^\text{srch}_\text{min}]$
\STATE If record $(u_{s},H_L,H_D,H_C)$ does not exist, sample $(u_s,H_L,H_D,H_C)\overset{\$}{\leftarrow}\{0\}\times\mathbb{G}\times\mathbb{G}\times\mathbb{G}$
\FOR{all $u^\prime\in\textsf{DUTime}(w)$ in ascending order where $u^\prime>u_s$}
\STATE  Retrieve $(u^\prime,L_0,D_0,C_0)$ from $\mathbf{CList}$
\STATE  Randomly select $tk\overset{\$}{\leftarrow}\{0,1\}^\lambda$
\STATE  Program random oracle $\mathbf{H}_1(tk)\leftarrow H_L$
\STATE  Program random oracle $\mathbf{H}_2(tk)\leftarrow H_D$
\STATE  Program random oracle $\mathbf{G}(tk)\leftarrow H_C$
\STATE  Set $(H_L,H_D)\leftarrow (L_0,\frac{D_0}{\pi(tk)})$
\STATE  Randomly sample $H_C\overset{\$}{\leftarrow}\mathbb{G}$
\ENDFOR
\STATE Retrieve the maximum timestamp $u^\text{DU}_\text{max}$ from $\textsf{DUTime}(w)$
\STATE Update $\mathbf{SHist}[u^\text{srch}_\text{min}]\leftarrow (u^\text{DU}_\text{max},H_L,H_D,H_C)$
\STATE Send the simulated trapdoor $(K_1,H_L^{K_1},H_D^{K_1}, H_C^{K_1})$ to the server via above secure channel
\STATE After receiving the results from the server, output all file identifiers extracted from $\textsf{TimeDB}(w)$
\end{algorithmic}

\leftline{\underline{$\mathcal{S}.\textsf{KeyUpdate}(\mathcal{L}^{KeyUpdt}_\text{Srv}=(\textsf{KUHist}(U_\text{now})))$}}
\begin{algorithmic}[1]
\STATE Establish a temporary secure channel with the server using the Diffie-Hellman key exchange protocol
\STATE Accumulate timestamp $u\leftarrow u+u$
\STATE Randomly draw the \textsf{KeyUpdate} token $\Delta$ from $\mathbb{Z}^*_q$
\STATE Update search key $K_1\leftarrow K_1\cdot\Delta$
\STATE Send $\Delta$ to the server via above secure channel
\end{algorithmic}
\end{multicols}
\end{algorithm*}

\textit{1) Proof of $\mathcal{L}_\text{Srv}$-Adaptive Security}
\begin{proof}
We construct a simulator $\mathcal{S}$ to prove that \texttt{Bamboo} is $\mathcal{L}_\text{Srv}$-adaptively secure. In $\mathcal{S}$, the hash functions $\mathbf{H}_1$, $\mathbf{H}_2$, and $\mathbf{G}$ are modeled as random oracles. Leakage functions $\mathcal{L}_\text{Srv}=(\mathcal{L}^{Stp}_\text{Srv},\mathcal{L}^{DaUpdt}_\text{Srv},$ $\mathcal{L}^{Srch}_\text{Srv},\mathcal{L}^{KeyUpdt}_\text{Srv})$ are defined as follows:
\begin{gather*}
\mathcal{L}^{Stp}_\text{Srv}(\lambda,a_\text{max})=(\lambda,a_\text{max}),~
\mathcal{L}^{DaUpdt}_\text{Srv}(op,(w,id))=\text{NULL},\\
\mathcal{L}^{KeyUpdt}_\text{Srv}=\mathcal{L}^\prime_\text{Srv}(\textsf{KUHist}(U_\text{now})),\\
\mathcal{L}^{Srch}_\text{Srv}(w)=\mathcal{L}^{\prime\prime}_\text{Srv}(\textsf{sp}(w),\textsf{TimeDB}(w),\qquad\qquad\qquad\quad\\
\qquad\qquad\qquad\qquad\qquad\qquad\textsf{DUTime}(w),\textsf{KUHist}(U_\text{now})),
\end{gather*}
where $\mathcal{L}^\prime_\text{Srv}$ and $\mathcal{L}^{\prime\prime}_\text{Srv}$ are two stateless functions.

The construction of $\mathcal{S}$ is shown in Algorithm~\ref{SIM.SEMI}.
In the simulated protocol \textsf{Setup}, $\mathcal{S}$ initializes a map $\mathbf{SHist}$, a list $\mathbf{CList}$, and a search key $K_1$.
The map $\mathbf{SHist}$ maintains the states of \textsf{Search} queries to guarantee the consistency of the search process.
The list $\mathbf{CList}$ records the ciphertexts produced by $\mathcal{S}$.
$\mathcal{S}$ will use $K_1$ to simulate the search trapdoor in protocol $\mathcal{S}.\textsf{Search}$.
Clearly, the simulated protocol $\textsf{Setup}$ is indistinguishable from \texttt{Bamboo}.\textsf{Setup}.

\begin{algorithm*}[h]
\algsetup{linenosize=\footnotesize} \footnotesize
\caption{Construction of Simulator $\mathcal{S}^\prime$.}\label{SIM.Out-1}
\begin{multicols}{2}
\leftline{\underline{$\mathcal{S}^\prime.\textsf{Setup}(\mathcal{L}^{Stp}_\text{Stl}(\lambda,a_\text{max})=(\lambda,a_\text{max}))$}}
\begin{algorithmic}[1]
\STATE Initialize an invertible mapping function $(1+l,\mathbb{G},q,\pi,$ $\pi^{-1})\leftarrow\mathbf{PGen}(\lambda,\ l+1)$
\STATE Initialize the Diffie-Hellman key exchange protocol parameters including the elliptic curve group $\mathbb{G}^\prime$ of prime order $q^\prime$ and a generator $g^\prime\in\mathbb{G}^\prime$
\STATE Initialize four empty maps $\mathbf{State}$, $\mathbf{TdMap}$, $\mathbf{OriCip}$, and $\mathbf{EDB}$
\STATE Initialize an empty list $\mathbf{CList}$
\STATE Initialize timestamp $u\leftarrow 0$
\STATE Initialize the search key $K_1\overset{\$}{\leftarrow} \mathbb{Z}^*_q$ and the encryption key $K_2\overset{\$}{\leftarrow} \mathbb{Z}^*_q$
\STATE Set the simulated key $K_\Sigma\leftarrow (K_1,K_2)$
\STATE Send the simulated database $\mathbf{EDB}$ to the server
\end{algorithmic}

\leftline{\underline{$\mathcal{S}^\prime.\textsf{DataUpdate}(\mathcal{L}^{DaUpdt}_\text{Stl}(op,(w,id))=\text{NULL})$}}
\leftline{Simulated Client:}
\begin{algorithmic}[1]
\STATE Accumulate timestamp $u\leftarrow u+1$ and insert $\mathbf{OriCip}[u]\leftarrow u$
\STATE Randomly select three elements $(L_0,D_0,C_0)\overset{\$}{\leftarrow}\mathbb{G}\times\mathbb{G}\times\mathbb{G}$
\STATE Compute the simulated ciphertext $L=L^{K_1}_0$, $D=D^{K_1}_0$, and $C=C^{K_1}_0$
\STATE Insert records $(0,u,L_0,D_0,C_0)$ and $(1,u,L,D,C)$ into $\mathbf{CList}$
\STATE Send the simulated ciphertext $(L,D,C)$ to the simulated server\\
\vspace{2pt}
\leftline{Simulated Server:}
\STATE Store the simulated ciphertext into the simulated database  $\mathbf{EDB}[L]\leftarrow (D,C)$
\end{algorithmic}

\leftline{\underline{$\mathcal{S}^\prime.\textsf{Search}(\mathcal{L}^{Srch}_\text{Stl}(w)=\text{NULL})$}}
\leftline{Simulated client:}
\begin{algorithmic}[1]
\STATE Establish a temporary secure channel with the simulated server using the Diffie-Hellman key exchange protocol
\STATE Accumulate timestamp $u\leftarrow u+1$
\STATE Randomly sample three elements $(e_1,e_2,e_3)\overset{\$}{\leftarrow}\mathbb{G}\times\mathbb{G}\times\mathbb{G}$
\STATE Send the simulated search trapdoor $(K_1,e_1,e_2,e_3)$ to the simulated server via above secure channel\\
\vspace{2pt}
\leftline{Simulated Server:}
\STATE Return $a_\text{max}$ random elements of $\mathbb{G}$ to the simulated client via above secure channel
\end{algorithmic}

\leftline{\underline{$\mathcal{S}^\prime.\textsf{KeyUpdate}(\mathcal{L}^{KeyUpdt}_\text{Stl}=\text{NULL})$}}
\leftline{Simulated Client:}
\begin{algorithmic}[1]
\STATE Establish a temporary secure channel with the server using the Diffie-Hellman key exchange protocol
\STATE Accumulate timestamp $u\leftarrow u+1$
\STATE Randomly draw the \textsf{KeyUpdate} token $\Delta$ from $\mathbb{Z}^*_q$
\STATE Update search key $K_1\leftarrow K_1\cdot\Delta$ and encryption key $K_2\leftarrow K_2\cdot\Delta$
\STATE Send $\Delta$ to the server via the above secure channel\\
\vspace{2pt}
\leftline{Simulated Server:}
\FOR{all ciphertext $(L,D,C)\textbf{ such that }(D,C)\leftarrow\mathbf{EDB}[L]$}
\STATE Accumulate $u\leftarrow u + 1$
\STATE Update ciphertexts $L^\prime\leftarrow L^{\Delta}$, $D^\prime\leftarrow D^\Delta$, and $C^\prime\leftarrow C^{\Delta}$
\STATE Insert ciphertext $\mathbf{EDB}[L^\prime]\leftarrow (D^\prime,C^\prime)$
\STATE Insert record $(1,u,L^\prime,D^\prime,C^\prime)$ into $\mathbf{CList}$
\STATE Retrieve $(1,u^\prime,L,D,C)$ from $\mathbf{CList}$
\STATE Retrieve $u_0\leftarrow\mathbf{OriCip}[u^\prime]$ and insert $\mathbf{OriCip}[u]\leftarrow u_0$
\STATE Remove $\mathbf{EDB}[L]$
\ENDFOR
\end{algorithmic}

\leftline{\underline{$\mathbf{S}^\prime.\textsf{KeyLeak}(\mathcal{L}^{KeyLeak}_\text{Stl}=(\textsf{CUHist}(U_\text{now}),\textsf{DUHist}(U_\text{now})))$}}
\begin{algorithmic}[1]
\STATE Accumulate timestamp $u\leftarrow u+1$
\STATE Initialize a list $\mathbf{LeakedData}\leftarrow\textsf{CUHist}(U_\text{now})\cup\textsf{DUHist}(U_\text{now})$
\FOR{each keyword $w$ existing in $\mathbf{LeakedData}$}
\STATE Initialize an empty list $\mathbf{DB}_w$
\FOR{each records $(u^\prime,op,(w,id))\in\mathbf{LeakedData}$}
\STATE Retrieve $u^\prime_0\leftarrow\mathbf{OriCip}[u^\prime]$ and insert $(u^\prime_0,op,id)$ into $\mathbf{DB}_w$
\ENDFOR
\FOR{all triplets $(u^\prime_0,op,id)\in\mathbf{DB}_w$ in the ascending order of $u^\prime_0$}
\STATE Retrieve search token $tk_w^\prime\leftarrow\mathbf{TdMap}[u^\prime_0]$
\STATE If $tk_w^\prime\neq\text{NULL}$, \textbf{continue}
\STATE Otherwise, randomly sample the search token $tk_w^\prime\overset{\$}{\leftarrow}\{0,1\}^\lambda$
\STATE Insert $\mathbf{TdMap}[u^\prime_0]\leftarrow tk_w^\prime$
\STATE Retrieve record $(0,u^\prime_0,L_0,D_0,C_0)$ from $\mathbf{CList}$
\STATE Retrieve record $(tk_w,cnt_w)$ from $\mathbf{State}[w]$
\IF{$(tk_w,cnt_w)=(\text{NULL},\text{NULL})$}
\STATE Randomly sample search token $tk_w\leftarrow \{0,1\}^\lambda$
\STATE Initialize \textsf{DataUpdate} counter $cnt_w\leftarrow 0$
\STATE Program random oracles $\mathbf{H}_1(tk_w)$,  $\mathbf{H}_2(tk_w)$, and $\mathbf{G}(tk_w)$ with randomly selected elements from $\mathbb{G}$
\ENDIF
\STATE Program random oracle $\mathbf{H}_1(tk_w^\prime)\leftarrow L_0$
\STATE Program random oracle $\mathbf{H}_2(tk_w^\prime)\leftarrow \frac{D_0}{\pi(tk_w)}$
\STATE Program random oracle $\mathbf{G}(tk_w^\prime)\leftarrow \frac{C_0}{\pi(op||id)^{\frac{K_2}{K_1}}}$
\STATE Update $\mathbf{State}[w]\leftarrow(tk_w^\prime,cnt_w+1)$
\ENDFOR
\ENDFOR
\STATE Expose $(K_1,K_2,\mathbf{State})$ to stealer $\mathcal{A}_\text{Stl}$
\end{algorithmic}
\end{multicols}
\end{algorithm*}

Protocol $\mathcal{S}.\textsf{DataUpdate}$ simulates the ciphertexts by sampling three random elements from $\mathbb{G}$ and computing the $K_1$ exponentiations of those random elements. Due to the one-wayness and collision resistance of $\mathbf{H}_1$, $\mathbf{H}_2$, and $\mathbf{G}$ and the randomness of the selected search token, a ciphertext $(L,D,C)$ generated by the real protocol \textsf{DataUpdate} is indistinguishable from random numbers before the client issues a \textsf{Search} query related to this ciphertext. Thus, this protocol is indistinguishable from the real  \texttt{Bamboo}.\textsf{DataUpdate}.

Protocol $\mathcal{S}.\textsf{Search}$ simulates \texttt{Bamboo}.\textsf{Search}.
$\mathcal{S}$ uses the minimum timestamp $u^\text{srch}_\text{min}$ of search pattern $\textsf{sp}(w)$ to indicate the latest simulated search trapdoor of $w$ in the map $\mathbf{SHist}$ (Steps 4 to 6).
The map $\mathbf{SHist}$ also records a timestamp $u_s$ to the newest simulated ciphertext whose corresponding search trapdoor has been programmed into random oracles $\mathbf{H}_1$, $\mathbf{H}_2$, and $\mathbf{G}$.
Any simulated ciphertext of $w$ of which the timestamp is older than $u_s$ will not be re-programmed into the oracles in the following steps.
In Steps 7 to 15, $\mathcal{S}$ randomly generates a search token for each simulated ciphertext and programs the search token into random oracles $\mathbf{H}_1$, $\mathbf{H}_2$, and $\mathbf{G}$.
The programming process follows the rule that for any two successive ciphertexts of the same keyword, the latter ciphertext encrypts the former ciphertext's search token. Under this rule, the server can correctly disclose the whole hidden chain of the same keyword with the search token of the latest simulated ciphertext of $w$.
The above guarantees that the server cannot distinguish the simulated search process from the real one. Besides,  $H_L^{K_1},H_D^{K_1},\text{and }H_C^{K_1}$ in the simulated search trapdoor are also indistinguishable from $L,Msk_D,\text{and }Msk_C$ of a real search trapdoor. Since the base numbers of $L,Msk_D,\text{and }Msk_C$ are hashed by randomly selected numbers (i.e., indistinguishable from random numbers), the simulated protocol \textsf{Search} is indistinguishable from the real one.

In terms of the simulated protocol \textsf{KeyUpdate}, it is clear that what the server $\mathcal{A}_\text{Srv}$ observes in $\mathcal{S}.\textsf{KeyUpdate}$ is identical to what $\mathcal{A}_\text{Srv}$ can observe in $\texttt{Bamboo}.\textsf{KeyUpdate}$. Thus, $\mathcal{S}$ simulates an indistinguishable protocol \textsf{KeyUpdate}.

To summarize, we construct an $\mathcal{S}$ that simulates an indistinguishable ideal game from the real game in view of the server with the given leakage functions $\mathcal{L}_\text{Srv}$. Hence, \texttt{Bamboo} is $\mathcal{L}_\text{Srv}$-adaptive where Leakage functions $\mathcal{L}_\text{Srv}=(\mathcal{L}^{Stp}_\text{Srv},\mathcal{L}^{DaUpdt}_\text{Srv},$ $\mathcal{L}^{Srch}_\text{Srv},\mathcal{L}^{KeyUpdt}_\text{Srv})$ are defined as:
\begin{gather*}
\mathcal{L}^{Stp}_\text{Srv}(\lambda,a_\text{max})=(\lambda,a_\text{max}),~
\mathcal{L}^{DaUpdt}_\text{Srv}(op,(w,id))=\text{NULL},\\
\mathcal{L}^{KeyUpdt}_\text{Srv}=\mathcal{L}^\prime_\text{Srv}(\textsf{KUHist}(U_\text{now})),\\
\mathcal{L}^{Srch}_\text{Srv}(w)=\mathcal{L}^{\prime\prime}_\text{Srv}(\textsf{sp}(w),\textsf{TimeDB}(w),\qquad\qquad\qquad\quad\\
\qquad\qquad\qquad\qquad\qquad\qquad\textsf{DUTime}(w),\textsf{KUHist}(U_\text{now})),
\end{gather*}
where $\mathcal{L}^\prime_\text{Srv}$ and $\mathcal{L}^{\prime\prime}_\text{Srv}$ are two stateless functions.
\end{proof}

\textit{2) Proof of $\mathcal{L}_\text{Stl}$-Adaptive Security}

\begin{proof}
We construct a simulator $\mathcal{S}^\prime$ to prove that \texttt{Bamboo} is $\mathcal{L}_\text{Stl}$-adaptively secure, where leakage functions $\mathcal{L}_\text{Stl}=(\mathcal{L}^{Stp}_\text{Stl},\mathcal{L}^{DaUpdt}_\text{Stl},$ $\mathcal{L}^{Srch}_\text{Stl},$ $\mathcal{L}^{KeyUpdt}_\text{Stl},\mathcal{L}^{KeyLeak}_\text{Stl})$ are defined as follows:
\begin{gather*}
\mathcal{L}^{Stp}_\text{Stl}(\lambda,a_\text{max})=(\lambda,a_\text{max}),~
\mathcal{L}^{DaUpdt}_\text{Stl}(op,(w,id))=\text{NULL},\\
\mathcal{L}^{Srch}_\text{Stl}(w)=\text{NULL},~
\mathcal{L}^{KeyUpdt}_\text{Stl}=\text{NULL},\\
\mathcal{L}^{KeyLeak}_\text{Stl}=\mathcal{L}^\prime_\text{Stl}(\textsf{CUHist}(U_\text{now}),\textsf{DUHist}(U_\text{now})),
\end{gather*}
where $\mathcal{L}^\prime_\text{Stl}$ is a stateless function.

Algorithm~\ref{SIM.Out-1} presents the construction of $\mathcal{S}^\prime$. In the construction of simulator $\mathcal{S}^\prime$, we utilize the DDH assumption and model hash functions $\mathbf{H}_1$, $\mathbf{H}_2$, and $\mathbf{G}$ as random oracles.

The simulated protocol \textsf{Setup} initializes the map $\mathbf{State}$ and two keys $K_1$ and $K_2$ to simulate the private state $\mathbf{State}$ and the secret key $K_\Sigma$, respectively.
$\mathcal{S}^\prime$ uses $K_\Sigma$ and $\mathbf{State}$ to simulate a key-compromise event. The list $\mathbf{CList}$ records the ciphertexts and their corresponding generation timestamp. The map $\mathbf{TdMap}$ records the search token used to generate ciphertexts. The map $\mathbf{OriCip}$ maps the timestamp of ciphertext to the timestamp of the original \textsf{DataUpdate} query that the former is key-updated from.
In the views of stealer $\mathcal{A}_\text{Stl}$, the simulation is indistinguishable from \texttt{Bamboo}.\textsf{Setup}.

In the simulated \textsf{DataUpdate}, $\mathcal{S}^\prime$ simulates the ciphertext by sampling three random numbers from $\mathbb{G}$ and computes the $K_1$ exponentiations of those elements (Steps 2 and 3).
In Step 4, $\mathcal{S}^\prime$ inserts the three elements and the simulated ciphertext along with the timestamps to the $\mathbf{CList}$.
The simulated ciphertext is indistinguishable from the real one via \texttt{Bamboo}.\textsf{DataUpdate} even if $\mathcal{A}_\text{Stl}$ compromises the secret key and the private state.
As a newly generated ciphertext is encrypted with a fresh random search token, the ciphertext is indistinguishable from a random number unless $\mathcal{A}_\text{Stl}$ holds the corresponding search token.
$\mathcal{A}_\text{Stl}$ cannot access the search token unless it obtains the secret key and private state again.

The simulated \textsf{Search} is indistinguishable from \texttt{Bamboo}.\textsf{Search}. With the protection of the secure channel, $\mathcal{A}_\text{Stl}$ cannot learn the search trapdoor and the search results.
The simulated client and server only need to select $a_\text{max}$ random data and transfer those data via the secure channel.

Similarly, the simulated protocol \textsf{KeyUpdate} is protected by the secure channel, i.e., $\mathcal{A}_\text{Stl}$ cannot get the \textsf{KeyUpdate} token.
Without the token, $\mathcal{A}_\text{Stl}$ cannot tell whether a post-key-updated ciphertext is updated from a pre-key-updated ciphertext even if $\mathcal{A}_\text{Stl}$ obtains \emph{one of} the secret keys of post-key-updated ciphertext and the pre-key-updated ciphertext.
This security property can be reduced to the DDH assumption in the group $\mathbb{G}$.
Specifically, suppose a searchable ciphertext $(L,D,C)$ can be written in the form of $(g^x,g^y,g^z)$, where $g$ is a generator of $\mathbb{G}$ and $x,y,z\in \mathbb{Z}^*_q$.
Notice that even if $\mathcal{A}_\text{Stl}$ learns the underlying secret key $(K_1,K_2)$, the entry $(op,(w,id))$, the search token $tk_w^\prime$, and the next search token $tk_w$ used to generate $(L,D,C)$, $\mathcal{A}_\text{Stl}$ cannot exactly know $x$, $y$, and $z$.
Specifically, suppose $\mathbf{H}_1(tk_w^\prime)=g^{h_1}$, $\mathbf{H}_2(tk_w^\prime)=g^{h_2}$, $\mathbf{G}(tk_w^\prime)=g^{h_3}$, $\pi(tk_w)=g^{n_1}$, and $\pi(op||id)=g^{n_2}$, we have $x=K_1\cdot h_1$, $y=(n_1+h_2)\cdot K_1$, and $z=K_2\cdot n_2 + K_1\cdot h_3$, wherein $h_1$, $h_2$, and $h_3$ are independent and random to $\mathcal{A}_\text{Stl}$.
Thus, $x$, $y$, and $z$ are independent and random to $\mathcal{A}_\text{Stl}$ due to the properties of the random oracles.
Suppose $\mathcal{S}^\prime$ holds the \textsf{KeyUpdate} token $\Delta\in \mathbb{Z}^*_q$, a random bit $b\leftarrow\{0,1\}$ and two different ciphertexts $(L_0,D_0,C_0)=(g^{x_0},g^{y_0},g^{z_0})$ and $(L_1,D_1,C_1)=(g^{x_1},g^{y_1},g^{z_1})$ where $x_0,y_0,z_0,x_1,y_1,z_1\in \mathbb{Z}^*_q$, and $\mathcal{A}_\text{Stl}$ is given $(g^{x_0},g^{y_0},g^{z_0})$, $(g^{x_1},g^{y_1},g^{z_1})$, and $(g^\Delta, g^{x_b\cdot \Delta}, g^{y_b\cdot \Delta}, g^{z_b\cdot \Delta})$ (note $\mathcal{A}_\text{Stl}$ actually never has access to $g^\Delta$ in both real and ideal games).
The advantage that $\mathcal{A}_\text{Stl}$ guesses a correct $b^*\in\{0,1\}$ such that $b^*=b$ is directly reduced to $\mathcal{A}_\text{Stl}$'s advantage to solve three DDH assumption instances, which is negligible.
Thus, the simulated \textsf{KeyUpdate} is indistinguishable from the real one in the view of $\mathcal{A}_\text{Stl}$.

When a key-compromise event happens, the function $\textsf{KeyLeak}$ will be called with the input of leaked information.
The goal of function $\textsf{KeyLeak}$ is to expose a secret key and a private state to $\mathcal{A}_\text{Stl}$ that allows $\mathcal{A}_\text{Stl}$ to correctly perform search over and decrypt the encrypted database $\mathbf{EDB}$.
$\mathcal{S}^\prime$ first extracts entries containing the same keywords $w$ from both $\textsf{CUHist}(U_\text{now})$ and $\textsf{DUHist}(U_\text{now})$, and stores those entries into $DB_w$ (Steps 4 to 7).
Secondly, for each entry in $DB_w$, $\mathcal{S}^\prime$ prepares the search token $tk^\prime$ that can be used to generate that entry.
If an entry has already been assigned a search token, $\mathcal{S}^\prime$ skips this entry and processes the successive one (Steps 8 to 10).
Then, $\mathcal{S}^\prime$ uses the information in $\mathbf{State}[w]$ to construct the hidden chain of $w$'s ciphertexts (Steps 14 to 23).
It programs the random oracles to construct the hidden chain of $w$'s ciphertext (Steps 20 to 22).
Finally, $\mathcal{S}^\prime$ exposes the secret key $K_\Sigma=(K_1,K_2)$ and private state $\mathbf{State}$ to $\mathcal{A}_\text{Stl}$.
After $\textsf{KeyLeak}$, the encrypted database contains real entries, and $\mathcal{A}_\text{Stl}$ can perform correct searches and decryption over the encrypted database with the exposed search key $K_1$ and encryption key $K_2$ and private state $\mathbf{State}$.
Thus, $\mathcal{A}_\text{Stl}$ cannot distinguish the simulated encrypted database from the real one.

To summarize, we construct an $\mathcal{S}^\prime$ that is indistinguishable from the real game in the views of $\mathcal{A}_\text{Stl}$. Thus, \texttt{Bamboo} is $\mathcal{S}_\text{Stl}$-adaptively secure where $\mathcal{L}_\text{Stl}=(\mathcal{L}^{Stp}_\text{Stl},\mathcal{L}^{DaUpdt}_\text{Stl},\mathcal{L}^{Srch}_\text{Stl},$ $\mathcal{L}^{KeyUpdt}_\text{Stl},\mathcal{L}^{KeyLeak}_\text{Stl})$ is defined as:
\begin{gather*}
\mathcal{L}^{Stp}_\text{Stl}(\lambda,a_\text{max})=(\lambda,a_\text{max}),~
\mathcal{L}^{DaUpdt}_\text{Stl}(op,(w,id))=\text{NULL},\\
\mathcal{L}^{Srch}_\text{Stl}(w)=\text{NULL},~
\mathcal{L}^{KeyUpdt}_\text{Stl}=\text{NULL},\\
\mathcal{L}^{KeyLeak}_\text{Stl}=\mathcal{L}^\prime_\text{Stl}(\textsf{CUHist}(U_\text{now}),\textsf{DUHist}(U_\text{now})),
\end{gather*}
where $\mathcal{L}^\prime_\text{Stl}$ is a stateless function.
\end{proof}

From the above security analysis, we prove that \texttt{Bamboo} is $\mathcal{L}_\text{Srv}$-adaptively and $\mathcal{L}_\text{Stl}$-adaptively secure with the leakage functions $\mathcal{L}_\text{Srv}$ and $\mathcal{L}_\text{Stl}$ defined in Theorem~\ref{THEO.SEKU-SEC}, respectively.

\end{document}